\documentclass[a4paper,11pt]{article}
\usepackage{jcappub} 
\usepackage{graphicx}
\usepackage{dcolumn}
\usepackage{bm}
\usepackage{hyperref}
\usepackage[utf8]{inputenc}
 \usepackage{textgreek}
\usepackage{multirow}
\usepackage{soul}
\usepackage{float}
\usepackage{graphicx}
\usepackage{array}
\usepackage{tabularx}
\usepackage{times}
\usepackage[normalem]{ulem}
\usepackage{color}
\usepackage{adjustbox}
\usepackage{cellspace}
\usepackage{cleveref}
\usepackage{booktabs}

\usepackage{lipsum}

\newcommand{\be}{\begin{equation}}
\newcommand{\ee}{\end{equation}}
\newcommand{\bea}{\begin{eqnarray}}
\newcommand{\eea}{\end{eqnarray}}

\newcommand{\comment}[1]{}
\renewcommand\sout{\bgroup \color{red} \ULdepth=-.5ex \ULset}
\def\simge{\mathrel{\rlap{\raise 0.511ex
     \hbox{$>$}}{\lower 0.511ex \hbox{$\sim$}}}}
\def\simle{\mathrel{\rlap{\raise 0.511ex
      \hbox{$<$}}{\lower 0.511ex \hbox{$\sim$}}}}

\usepackage{xcolor}

\arxivnumber{2510.08115} 
\title{\boldmath Probing Strange
Dark Matter through $f$-mode Oscillations of Neutron Stars with Hyperons and Quark Matter}

\author[a]{Mahboubeh Shahrbaf,\ }

\emailAdd{mahboubeh.shahrbafmotlagh@uwr.edu.pl}

\author[b]{Prashant Thakur,\ }

\author[a]{Davood Rafiei Karkevandi}

\affiliation[a]{Institute of Theoretical Physics, University of Wroclaw\\
Plac Maxa Borna 9, 50-204 Wroclaw, Poland}
\affiliation[b]{Department of Physics, Yonsei University\\
Seoul, 03722, South Korea}

\abstract{We investigate the impact of a hypothetical bosonic dark matter (DM) candidate, the sexaquark, on the fundamental ($f$-mode) oscillations of neutron stars (NSs). By varying the DM particle mass and considering different core compositions including hypernuclear matter, sexaquark DM, and deconfined quark matter (QM), we construct hybrid equations of state (EOS) with a smooth hadron--quark crossover that remain consistent with current astrophysical constraints on mass {($M$)}, radius {($R$)}, and tidal deformability {($\Lambda$)}. Our analysis shows that the presence of these exotic components systematically alters quasi-universal $f$-mode relations {considering $f$-mode frequency ($f$), damping time ($\tau$),  compactness  ($C$), and angular velocity ($\omega$)}. In particular, relations involving $f$--$\sqrt{M/R^{3}}$, $(R^{4}/M^{3}\tau)(C)$, $\omega M(C)$, require higher-order polynomial fits compared to standard studies. Quadratic forms remain sufficient for $f$--$\sqrt{M/R^{3}}$ and $\omega M(C)$, while damping-time relations such as $(R^{4}/M^{3}\tau)(C)$ demand higher-order corrections to capture their curvature. For $f(\Lambda)$, a cubic fit provides a satisfactory description. Within this extended framework the relations remain tight and effectively composition independent. These results suggest that precise $f$-mode measurements with future gravitational-wave detectors could provide clear signatures of DM and other exotic matter in NS interiors.}

\begin{document}
\maketitle
\flushbottom

\section{Introduction}
\label{sec:intro}

{Neutron stars (NSs), thanks to their high density and extreme gravitational potential, are one of the promising astrophysical laboratories for probing the nature of dark matter (DM) \cite{Baryakhtar:2022hbu,Bramante:2023djs,Grippa:2024ach}. NSs can accumulate DM through various mechanisms, including accretion in different stages of the NS life or production scenarios \cite{Ellis:2018bkr,Nelson:2018xtr,Ivanytskyi:2019wxd}.

The presence of DM in the NS can alter the mass-radius profile, tidal deformability, and oscillation modes, offering indirect probes of DM properties through gravitational and electromagnetic observations \cite{Karkevandi:2021ygv,Shakeri:2022dwg,Thakur:2023aqm,Thakur:2024btu,Shahrbaf:2024gdm,Thakur:2025zhi}.}
Moreover, DM can influence the thermal evolution of both cold NS \cite{Sedrakian:2015krq,Avila:2023rzj} and proto-NS   \cite{Issifu:2024htq}, and it can impact rotational dynamics, particularly in rapidly rotating stars  \cite{Konstantinou:2024ynd,Shawqi:2025cca,Issifu:2025gsq}. In binary neutron star (BNS) mergers, a DM component can imprint the gravitational wave (GW) signal during the inspiral phase and shift characteristic post-merger spectral peaks \cite{Giangrandi:2025rko}. Together, these effects could offer explanations for certain astrophysical observations that deviate from predictions based on standard NS models.

Recent multi-messenger observations provide powerful constraints on the NS {equation of state} (EOS), which can in turn be used to test the viability of models involving DM admixed NSs. For instance, precise measurements of the maximum mass of NSs, such as PSRJ 0740+6620 with a mass of approximately $2.08~M_\odot$ \cite{NANOGrav:2019jur}, and tidal deformability constraints from the binary NS merger GW170817 \cite{LIGOScientific:2018cki} restrict the range of viable EOSs and, consequently, the parameter space of DM models that can coexist with these observations \cite{Rutherford:2022xeb,Shakeri:2022dwg,Shahrbaf:2024gdm,Thakur:2025oyn}. Furthermore, future detections of NS oscillations, especially the fundamental modes, may reveal characteristic imprints of DM in the core, offering a novel, complementary window into dark sector physics through GW asteroseismology \cite{Shirke:2023ktu,Celato:2025klx,Shirke:2025ust,Flores:2024hts,Dey:2024vsw,Thakur:2025zhi}. 

Isolated NSs exhibit a spectrum of oscillation modes governed by various restoring forces, each offering a distinct probe into their dense interior. Of particular interest are the non-radial modes, which can generate time-varying mass quadrupole moments and thus serve as potent sources of  GWs. These oscillations can be excited by a range of astrophysical phenomena, including BNS mergers, magnetar flares, and starquakes~\cite{Ferrari:2002ut,Mock_1998, Kokkotas:1999mn,LIGOScientific:2019ccu, LIGOScientific:2022sts}. The gravitational radiation emitted during such events encodes valuable information about the star's internal composition and EOS, enabling a unique observational avenue to explore the physics of ultra-dense matter beyond the reach of terrestrial laboratories.

Non-radial oscillation modes are typically categorized by their parity. Polar modes include the fundamental ($f$), gravity ($g$), and pressure ($p$) modes, while axial modes encompass the rotational ($r$) and space-time ($w$) modes~\cite{Kokkotas:1999bd,Benhar:2004xg}. Among these, the $f$-mode stands out due to its strong coupling to gravitational radiation and its close connection with the tidal deformability $\Lambda$ of NSs, an observable that plays a critical role during the inspiral phase of BNS coalescences. The detection of $f$-mode signals, therefore, holds great promise for constraining the EOS, especially with the advent of current and next-generation GW observatories such as Advanced LIGO, Virgo, KAGRA, the Einstein Telescope (ET), and the Cosmic Explorer (CE)~\cite{Punturo_2010,PhysRevLett.122.061104,Pratten:2019sed,Pradhan:2023zor,Thakur:2025qwl,Sotani:2021nlx,Sotani:2010mx}.

The $f$-mode, being the fundamental polar mode with no radial nodes ($n = 0$), typically spans frequencies between $1.3$ and $2.8$\,kHz~\cite{Kunjipurayil:2022zah}. In contrast, $p$-modes, dominated by pressure as the restoring force, and $w$-modes, governed by the curvature of spacetime itself, appear at significantly higher frequencies ($5$--$12$\,kHz) and are generally more difficult to excite in astrophysical settings due to their strong damping~\cite{Kokkotas:1999bd}. Given these considerations, the present work focuses on the $f$-mode as the most astrophysically relevant oscillation for probing NS interiors and their GW signatures.

The EOS of NS matter plays a central role in determining the star's internal composition and macroscopic properties. To construct the EOS, researchers employ a range of theoretical frameworks which account for the interactions among nucleons, hyperons, and potentially other exotic constituents such as deconfined quark matter \cite{Baldo:2003vx, Shahrbaf:2019wex, Shahrbaf:2019bef, Shahrbaf:2019vtf, Tolos:2020aln, Oertel:2016bki,Looee:2025dgx,Sedrakian:2022ata,Tsiopelas:2024ksy}. These models are constrained and calibrated through a combination of experimental data and astrophysical observations \cite{Chatziioannou:2024tjq,Koehn:2024set,Kochankovski:2025lqc}. {In this regard, the extreme densities inside NSs can lead to DM production, resulting in a non-negligible dark component that alters the EOS and, consequently, the star's structure, stability, and macroscopic observables \cite{Shahrbaf:2024gdm,Thakur:2024btu}.}

Recent investigations have deepened our understanding of the microphysical structure and macroscopic characteristics of NSs by considering the role of DM in NS interiors \cite{ Thakur:2024btu, Karkevandi:2024vov, Biesdorf:2024dor, Pitz:2024xvh, Sagun:2023rzp,Rutherford:2024uix}. Numerous studies have aimed to constrain the properties of DM particles using observational limits from NSs, employing both single-fluid \cite{Hajkarim:2024ecp,Lopes:2024ixl,Shahrbaf:2023uxy} and two-fluid frameworks \cite{Karkevandi:2021ygv, RafieiKarkevandi:2021hcc, Thakur:2023aqm}. In particular, several studies have explored how the inclusion of DM within NSs can influence their oscillation spectra, focusing on the $f$-mode and p-mode \cite{Kain:2021hpk,Shirke:2024ymc,Sen:2024yim,Thakur:2025zhi,Jyothilakshmi:2024xtl,Caballero:2024qtv,Shirke:2025lsu,Routaray:2025gbq}. These oscillations, which are sensitive to the star's internal composition and structure, may carry distinctive signatures of DM, offering a potential probe for its presence in compact stars. It is worth mentioning that some of the aforementioned works need to be refined by incorporating full General Relativity (GR) calculations since they are limited to the Cowling approximation. The Cowling approximation, which neglects perturbations to the gravitational field, simplifies the analysis but can result in less accurate predictions, especially for the dynamics and oscillations of compact objects like NSs. The inclusion of full GR ensures a more comprehensive and precise treatment of both the gravitational and matter fields, offering a more reliable framework for studying the structure and behavior of NSs, particularly in the context of exotic matter \cite{Kruger:2024fxn,Rather:2024mtd,Sen:2025ndg,Pradhan:2022vdf}.

To explore the mixed DM scenario in NS, it is essential to understand the EOS for the DM admixed NS. We adopt a single-fluid DM-admixed framework in which DM is produced inside the NS. In this work, the EOS of hadronic matter is described using a relativistic mean-field (RMF) model, which incorporate many-body interactions through meson exchange called DD2Y-T model \cite{Typel:2009sy}. The model includes the $\sigma$, $\omega$, $\rho$, and $\Phi$ mesons, with the coupling between octet baryons (nucleons and hyperons) and mesons being density-dependent. We consider the double-strangeness bosonic sexaquark (S) as our DM candidate, a spin-color-flavor singlet composed of six quarks (uuddss), initially proposed by Farrar and collaborators \cite{Farrar:2017eqq, Farrar:2022mih}. Although the existence of S particles remains hypothetical, their unique properties and potential role as a stable DM candidate have attracted growing interest in recent years \cite{She:2025dqx, Gal:2024nbr, Moore:2024mot, Evans:2023zde, Doser:2023gls, Farrar:2023wvm}. In our study, the S particles are incorporated as an additional degree of freedom in the EOS of hadronic matter, enabling it to influence the NS’s internal composition and macroscopic observables, such as mass, radius, and tidal deformability \cite{Shahrbaf:2022upc, Shahrbaf:2024gdm, Shahrbaf:2023uxy}.

In this study, we perform fully general‐relativistic calculations of the $f$-mode oscillation frequencies and their corresponding damping times for DM admixed hybrid stars. Our stellar model is constructed via a smooth crossover transition from a hadronic phase, comprising nucleons, hyperons, and the bosonic S DM, to a color-superconducting quark phase described within the nonlocal Nambu-Janao-Lasinio (nlNJL) model \cite{Blaschke:2007ri}. By embedding all these particle species (including DM) within a single hybrid EOS, we employ a comprehensive framework to investigate the oscillation behavior of such compact objects. Indeed, the influence of each additional degree of freedom, with particular emphasis on strangeness-bearing DM and its competition with hyperons, has been systematically examined. In addition, we analyze 
$f$-mode quasi-universal relations and test their robustness in the presence of DM, enabling EOS-insensitive inference for DM-admixed hybrid stars. These relations offer a practical way for interpreting measurements of GW frequencies and amplitudes. Despite existing EOS uncertainties, they could enable the direct inference of stellar properties such as mass, radius, and tidal deformability from measured $f$-mode frequencies and damping times in future GW detections.

The rest of this paper is organized as follows. In Sec. \ref{sec:formalism}, we describe our approach for formulating the hadronic and quark matter EOSs for NSs, as well as the construction of the phase transition between the two phases. Sec. \ref{sec:GRR} provides the theoretical background for $f$-mode oscillations. The results and discussion, based on multi-messenger observational constraints of NSs, stellar structure profiles, and $f$-mode frequencies, are presented in Section \ref{sec:results}. Section \ref{sec:universal} presents the universal relation for DM admixed hybrid stars containing hyperons.
Finally, in Sec. \ref{sec:summary}, we summarize our findings and present our conclusions.

\section{Model Description and Formalism}
\label{sec:formalism}
In this work, bosonic DM particles (S) are produced in the hadronic sector through non-gravitational interactions with the baryons of the medium. The EOS for hadronic matter is derived using a generalized relativistic density functional (GRDF), where the meson-baryon couplings depend on the baryon density \cite{Typel:1999yq, Typel:2005ba}. The parametrization used for these couplings is known as DD2, and its density dependence is tuned to reproduce the properties of atomic nuclei \cite{Typel:2009sy}. The hadronic phase, which includes all hyperons in addition to nucleons and strange DM, is described within the DD2Y-T framework.

To prevent the immediate collapse of the star due to the Bose-Einstein condensation of bosonic S particles, we introduced a positive density-dependent mass shift for these particles. This mass shift also effectively accounts for interactions between baryonic matter and DM. As a result, all baryons in $\beta$-equilibrated NS matter acquire an effective mass ($m_i^* = m_i - S_i$) and an effective chemical potential ($\mu^*_i = \mu_i - V_i$), where $S_i$ and $V_i$ represent the scalar and vector potentials, respectively. The scalar potential for the S particles corresponds to their mass shift; thus, their effective mass reads
\begin{equation}
    m^*_S = m_S - S_S = m_S + \Delta m_S = m_S (1+x \frac{n_b}{n_0})
    \label{Smass}
\end{equation}
where we assumed the mass shift of S increases linearly with the baryon density for our initial exploration \cite{Shahrbaf:2022upc}. Here, the vacuum mass of the S particle ($m_S$), the baryon density ($n_b$), and the nuclear saturation density ($n_0 = 0.16 fm^{-3}$) are key parameters. 
{The parameter $x$, which controls the density dependence of the S mass, is varied in the range $0.03 \leq x \leq 0.10$. This interval is directly determined from the scan of the $(m_S,x)$ parameter space performed in \cite{Shahrbaf:2022upc}, where the constraints from nuclear physics, NS, and cosmological stability (summarized in Fig.~8 of that paper) restrict the phenomenologically viable values of $x$ for the DM motivated mass range of the S to approximately this interval.} This parameter represents the positive slope of the mass shift and quantifies the coupling strength of DM to baryons.

Recent investigations into NSs and dense nuclear matter indicate that quantum chromodynamics (QCD) matter undergoes a rapid stiffening above saturation density, progressing more strongly than in purely hadronic descriptions \citep{Kojo:2024ejq}. While unambiguous observational evidence for deconfined QM in a realistic EOS is still lacking, theoretical studies have demonstrated that post-merger GWs could provide a discriminating signature between scenarios with and without deconfinement \citep{Fujimoto:2022xhv}. Furthermore, several works have argued that GW observations, together with signals from core-collapse supernovae, NS mergers, and accreting NSs, are likely to be influenced by the occurrence of a phase transition to QM \citep{Bauswein:2018bma, Most:2018eaw, Weih:2019xvw, Bauswein:2022vtq, Largani:2023kjx}. In addition, a peak in the squared speed of sound of QCD matter has been linked to the onset of the phase transition to deconfined quark matter \cite{Satz:1998kg, Magas:2003wi, Castorina:2008vu, Braun-Munzinger:2014lba, Fukushima:2020cmk, Marczenko:2022jhl}. Taken together, these results strongly suggest that a transition to deconfined QM in the interior of compact stars may be unavoidable. 

Thus, it is reasonable to consider that the core of NS may contain not only hadronic matter but also QM. For a comprehensive analysis that accounts for all main possible degrees of freedom in the NS core, and to compensate for the low maximum mass of pure hadronic star when the parameter $x$ is small \cite{Shahrbaf:2022upc}, we consider a phase transition to deconfined QM at high densities. The EOS for quark sector is modeled using nlNJL approach \cite{Blaschke:2007ri}, which accounts for quark color superconductivity. For computational convenience, it is fitted to the constant speed of sound (CSS) parameterization \cite{Shahrbaf:2023uxy}.

For constructing a phase transition to QM, traditional Maxwell constructions typically lead to a first-order phase transition. However, a continuous transition, known as a crossover in the QCD context, is theoretically supported \citep{Schafer:1998ef, Hirono:2018fjr, Fujimoto:2019sxg}. Percolation theory indicates that the transition from hadronic to QM may proceed as a smooth crossover, consistent with lattice QCD results at high temperatures, and offers insight into hadron-quark continuity at baryon densities of $2n_0 < n_b < (4-7)n_0$ \cite{Fukushima:2020cmk, Baym:2017whm}.
Moreover, observations of similar radii for $1.4$ and $2.1M_\odot$ NSs disfavor a strong first-order phase transition up to $n_b \approx (4-7)n_0$, supporting hadron–quark continuity \cite{Kojo:2020krb, Kojo:2021hqh}.


For modeling the smooth phase transition, we employ the Replacement Interpolation Construction (RIC) method \cite{Ayriyan:2021prr, Ayriyan:2017nby, Ayriyan:2017tvl}. In this approach, a second-order polynomial interpolates between the hadronic and quark phases, producing a smooth crossover. The interpolated region contains both hadronic and QM, hence it is treated as a mixed-phase construction. To maintain thermodynamic consistency, the interpolation is performed at the level of $P(\mu_B)$, with all other thermodynamic quantities obtained by differentiation with respect to $\mu_B$ (for details on the RIC method, see \cite{Shahrbaf:2024gdm}).

\begin{figure}[htbp]
\centering
	\includegraphics[width=0.7\textwidth]{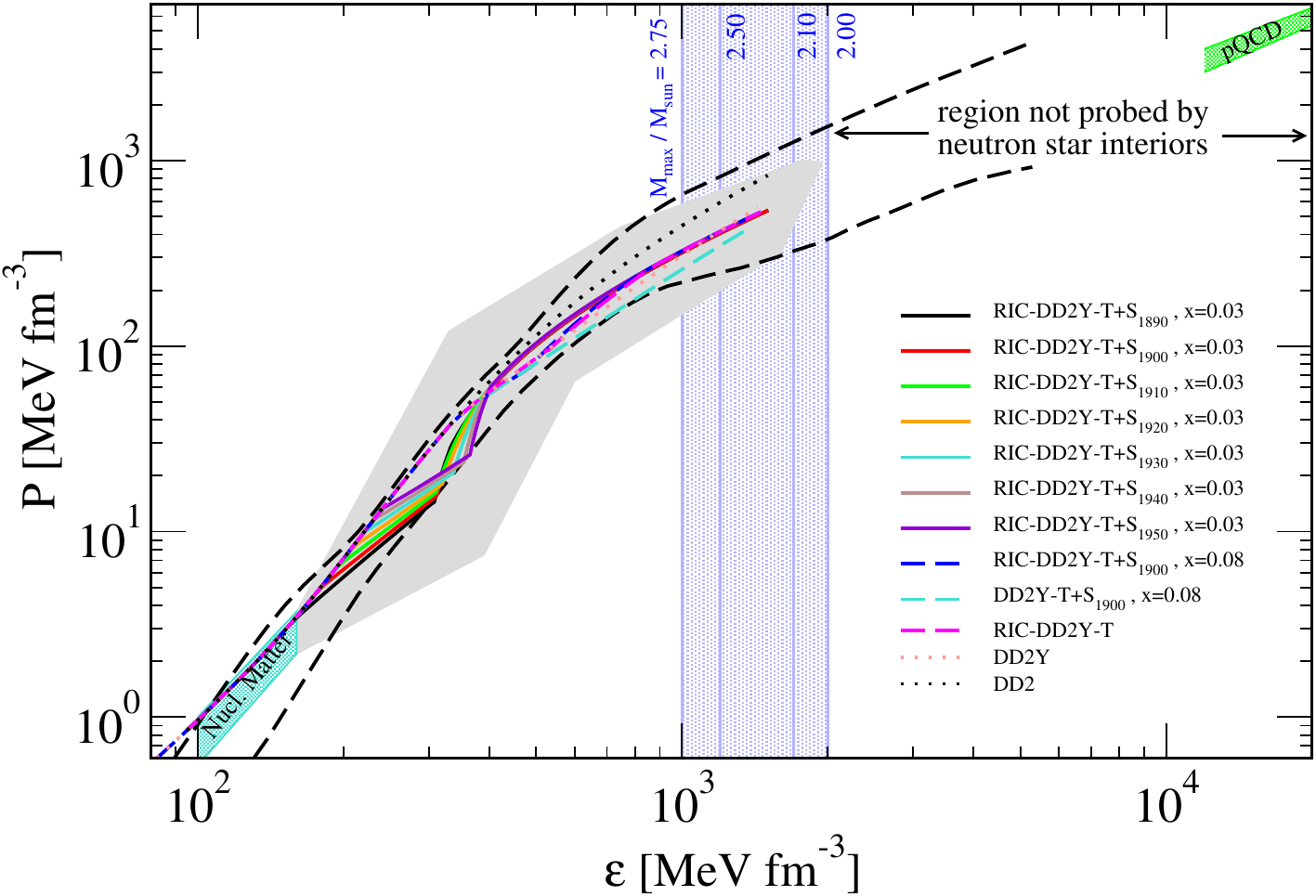}

	\caption{Pressure as a function of energy density for all EOSs. Pure hadronic models with  (DD2Y-T+S) and without (DD2 and DD2Y-T) DM, as well as hybrid models featuring a quark matter core (RIC-DD2Y-T and RIC-DD2Y-T+S), are shown. Various DM masses and two different values of the coupling parameter $x$ are used to explore different scenarios in this work. The constraints from Hebeler et al. \cite{Hebeler:2013nza} and Miller et al. \cite{Miller:2021qha} are represented by the gray region and the black dashed line, respectively. 
		\label{fig:p-e}
	}
\end{figure}

{Figure~\ref{fig:p-e} displays the EOS in the pressure-energy density plane for all models considered in this study. The plot includes the DD2, which is a purely nucleonic model, and DD2Y-T, which incorporates hyperons. The EOS labeled DD2Y-T+S stands for the hadronic model including DM in the form of S particles, in addition to nucleons and hyperons. The curves labeled RIC-DD2Y-T+S and RIC-DD2Y-T represent hybrid EOSs constructed via a crossover transition from the hadronic phase to deconfined QM, with and without S DM, respectively. Models that include DM are distinguished by the assumed mass of the S particle and the value of the slope parameter $x$, which characterizes the coupling strength between the S particle and baryons. Astrophysical constraints from Hebeler et al. \cite{Hebeler:2013nza} and Miller et al. \cite{Miller:2021qha} are shown as the gray region and black dashed lines, respectively. {All EOSs lie well within the broader gray constraint region and additionally pass through the most restrictive constraint band defined by the black dashed lines.} This result is consistent with our previous work \cite{Shahrbaf:2024gdm}, in which we first showed that the favored S masses, lying within the $68\%$ posterior interval, are in the range $m_S = 1890$-$1935$ MeV based on Bayesian analysis. Second, we found that the only S mass range fully consistent with all current observational constraints is the narrower range $m_S = 1890$-$1900$ MeV. Therefore, S masses above $1950$~MeV {are strongly disfavored} and the main analysis in this study focuses on S DM masses in the range $m_S = 1890$-$1950$~MeV.
 The vertical blue lines indicate the maximum central energy densities reached in the cores of NSs with varying masses (in units of solar mass). Clearly, none of our EOSs exceed these limits while they all respect the nuclear matter constraints at low densities.

{Furthermore, with regard to thermodynamic stability and causality, the monotonic increase of the pressure with energy density shown in Fig.~\ref{fig:p-e} demonstrates that all considered EOSs satisfy the fundamental stability conditions. In addition, for the range of S masses considered in this work ($m_S = 1890$-$1950$ MeV), we have explicitly verified that these conditions are fulfilled for the entire EOS set.
 This was done by computing the adiabatic sound speed
$c_s^2=dP/d\varepsilon$ and confirming $0 \le c_s^2 \le 1$ over the full density range reached by our stellar models
(see Fig.~\ref{fig:cs2}).
}

\begin{figure}[htbp]
\centering
	\includegraphics[width=0.7\textwidth]{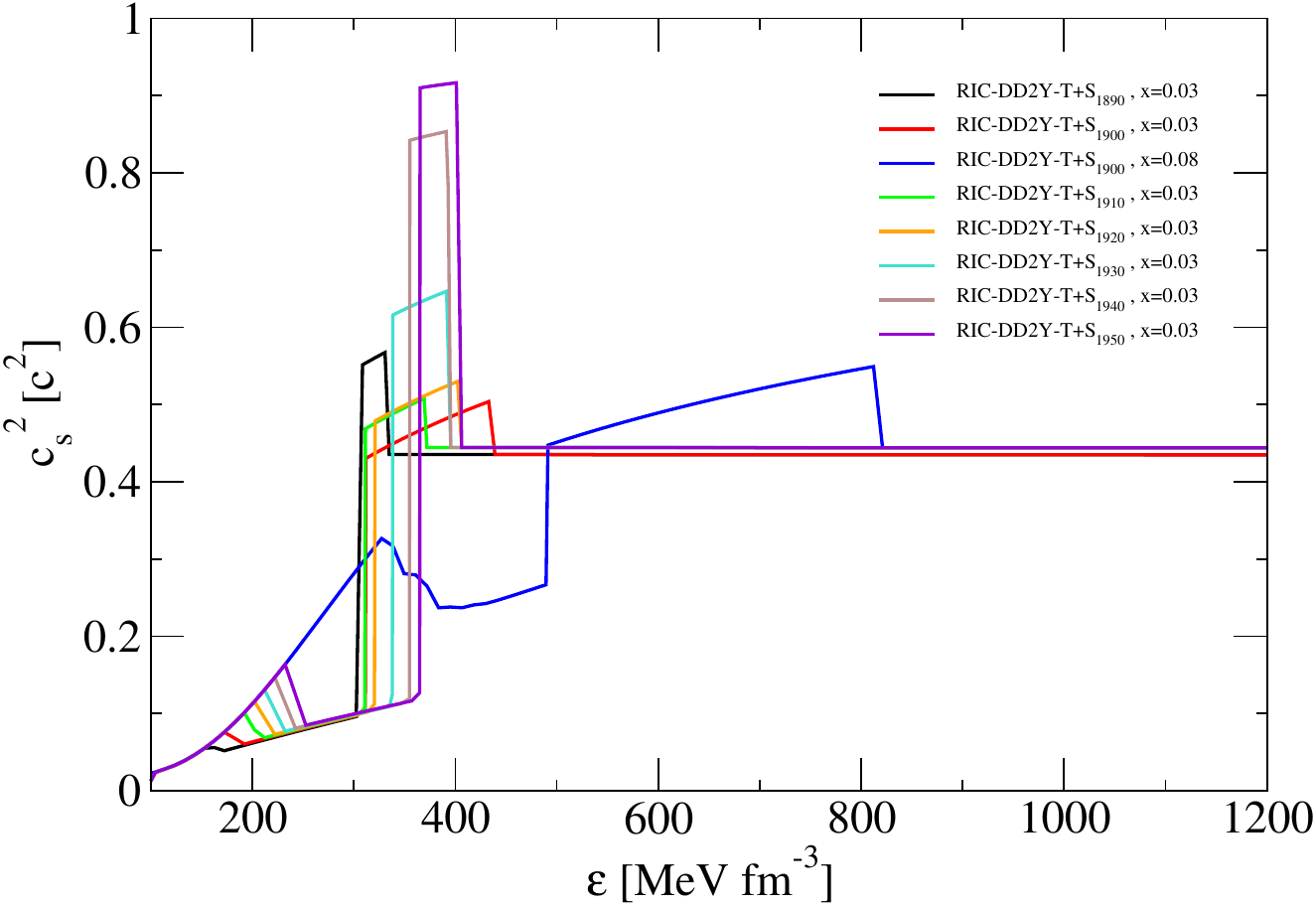}

	\caption{{The squared speed of sound as a function of energy density for the EOSs relevant to the current study. 
    }
		\label{fig:cs2}
	}
\end{figure}

In this work, we explore three different scenarios to investigate the impact of key parameters on NS observables and $f$-mode oscillations. These scenarios are as follows:

\begin{itemize} 
\item First Scenario: Hybrid EOSs featuring a quark core surrounded by hadronic matter composed of nucleons, hyperons, and S DM. These EOSs are labeled as RIC-DD2Y-T+S and correspond to different masses of S particles, while keeping the coupling parameter fixed at $x=0.03$. This scenario aims to examine the effect of varying DM mass on NS observables and $f$-mode frequencies.

\item Second Scenario: Hybrid EOSs with and without DM, both containing a quark core and hadronic matter made of nucleons and hyperons. The corresponding models are labeled as RIC-DD2Y-T+S (with DM) and RIC-DD2Y-T (without DM). The mass of S particles is set to $m_S=1900$ MeV, 
and $x=0.03$. This scenario is designed to assess the role of DM in hybrid stars.

\item Third Scenario: A hybrid EOS with a quark core and hadronic matter, including nucleons, hyperons, and S DM is compared to three pure hadronic EOSs: DD2 (nucleons only), DD2Y-T (nucleons + hyperons), and DD2Y-T+S (nucleons + hyperons + DM). The mass of the S DM is fixed at 
$1900$ MeV, while the coupling parameter is increased to $x=0.08$ to satisfy the maximum mass constraint from NS observations in the pure hadronic case with DM. This scenario investigates how the internal structure of a compact star and its observational values are affected by the inclusion of additional degrees of freedom - namely hyperons, DM, and deconfined QM. 
\end{itemize}

\section{Non-radial oscillations}
\label{sec:GRR}
Within the full GR framework, non-radial oscillations of NSs are studied by introducing small perturbations to the static background spacetime metric. These perturbations give rise to oscillation modes whose complex eigenfrequencies contain both the oscillatory behavior (real part) and the damping due to GW emission (imaginary part). In this formulation, the complete set of Einstein’s field equations is solved by treating GWs as perturbations of the static metric of a non-rotating star. In contrast, the Cowling approximation simplifies the analysis by neglecting perturbations of the gravitational field and considering only fluid perturbations inside the star. This omission of the back-reaction on the gravitational potential significantly reduces the computational complexity but leads to less accurate frequency estimates, as it does not fully account for the coupled dynamics between the metric and the fluid that are intrinsic to the full GR treatment~\cite{Pradhan:2022vdf,Kunjipurayil:2022zah,Roy:2023gzi}. In the present work, we employ the full GR framework to compute the $f$-mode frequencies. For a detailed derivation and the governing equations, we refer the reader to Ref.~\cite{Rather:2024mtd}.

\section{Results and Discussion}
\label{sec:results}

\subsection{Multi-messenger observational constraints of NSs}


\begin{figure*}[htbp]
  \centering
  \begin{minipage}[b]{0.32\linewidth}
    \centering
    \includegraphics[width=\linewidth]{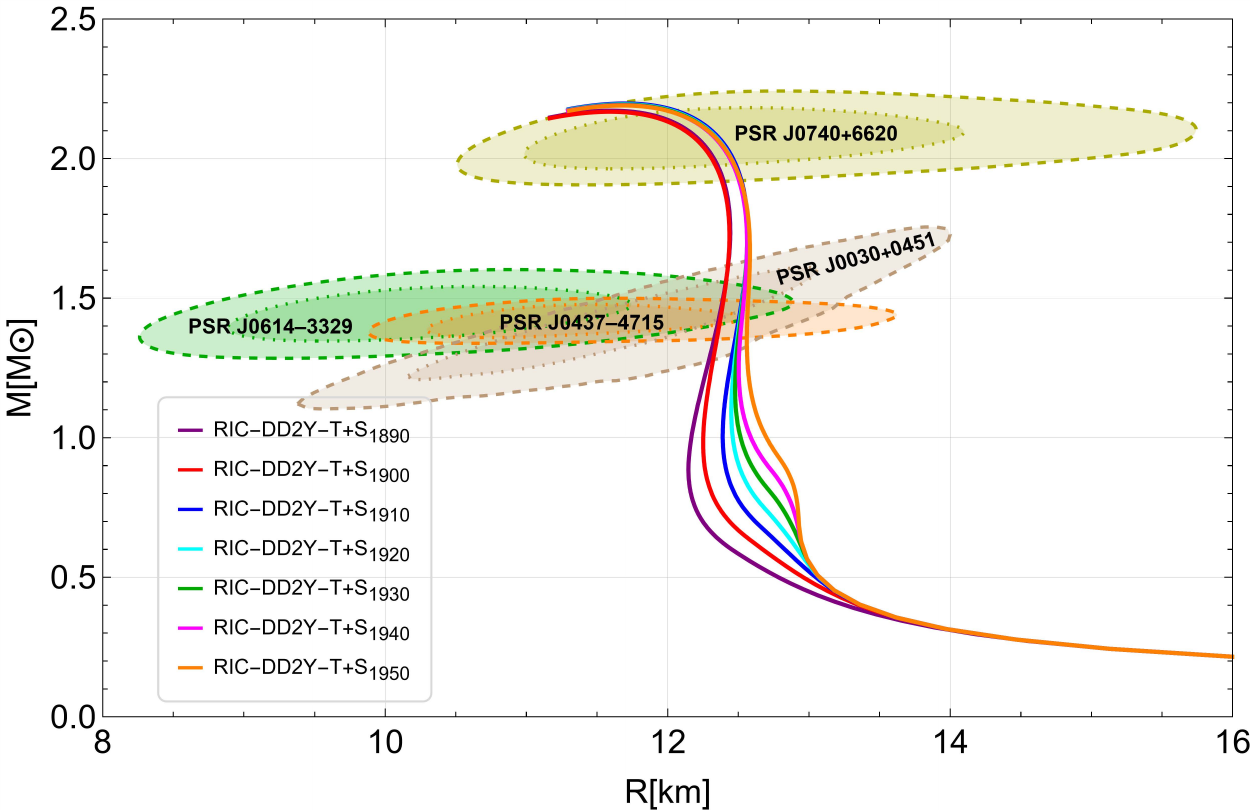}
    \par\medskip
    \centering\textbf{First scenario}
  \end{minipage}\hfill
  \begin{minipage}[b]{0.32\linewidth}
    \centering
    \includegraphics[width=\linewidth]{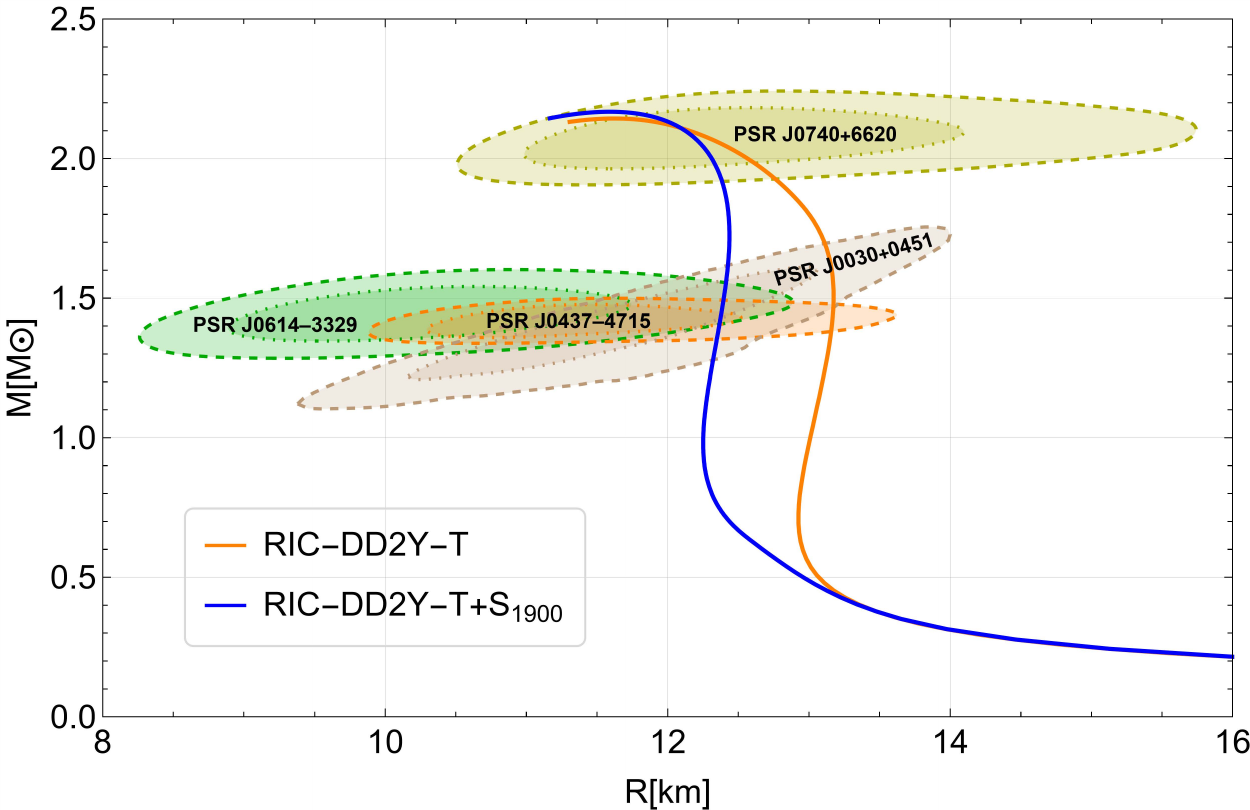}
    \par\medskip
    \centering\textbf{Second scenario}
  \end{minipage}\hfill
  \begin{minipage}[b]{0.32\linewidth}
    \centering
    \includegraphics[width=\linewidth]{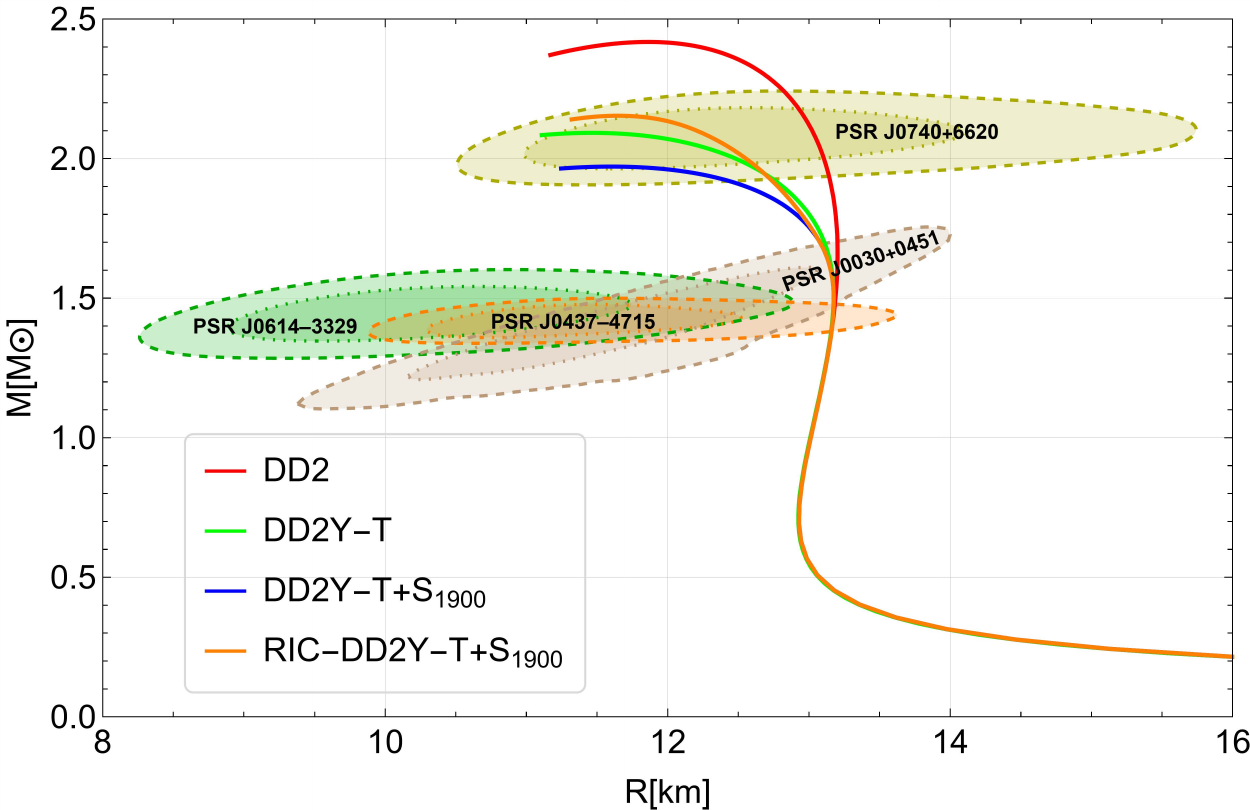}
    \par\medskip
    \centering\textbf{Third scenario}
  \end{minipage}
  \caption{Mass-Radius profiles for three scenarios considered in this work. The left panel indicates the effect of different masses of S (DM) particles ($x=0.03$), the middle one shows the impacts of DM emergence on hybrid stars ($x=0.03$), while the right one depicts how each new degree of freedom affects the mass and radius of the object ($x=0.08$). The colored regions represent observational constraints from the NICER mass-radius measurements.}
  \label{fig:M-R}
\end{figure*}


\begin{figure*}[htbp]
  \centering
  \begin{minipage}[b]{0.32\linewidth}
    \centering
    \includegraphics[width=\linewidth]{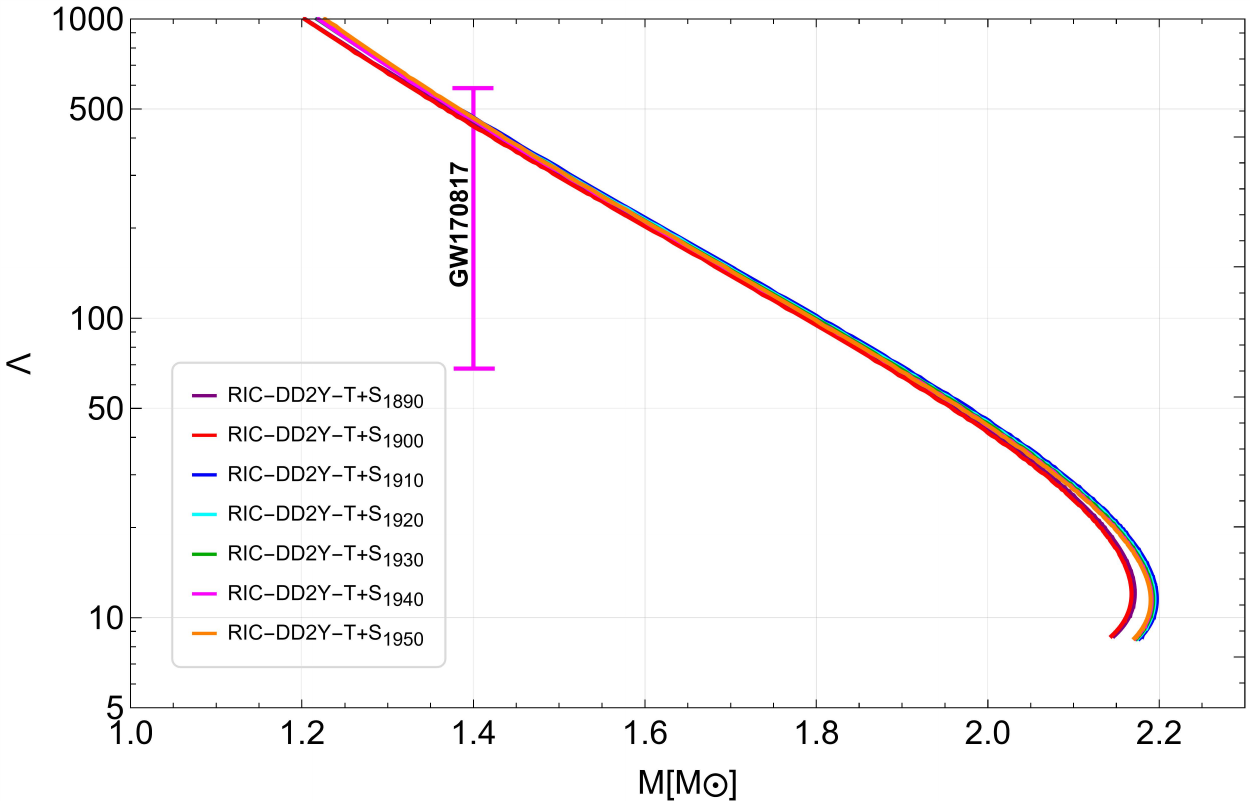}
    \par\medskip
    \centering\textbf{First scenario}
  \end{minipage}\hfill
  \begin{minipage}[b]{0.32\linewidth}
    \centering
    \includegraphics[width=\linewidth]{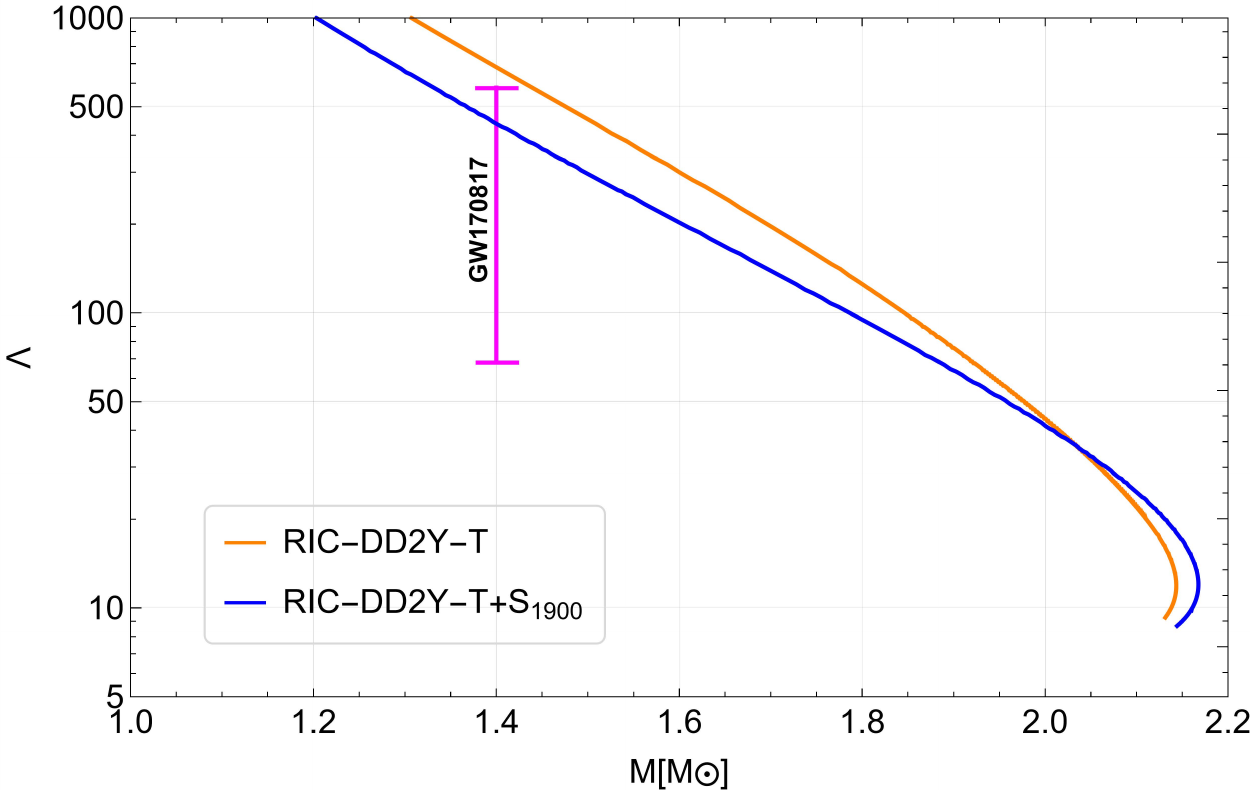}
    \par\medskip
    \centering\textbf{Second scenario}
  \end{minipage}\hfill
  \begin{minipage}[b]{0.32\linewidth}
    \centering
    \includegraphics[width=\linewidth]{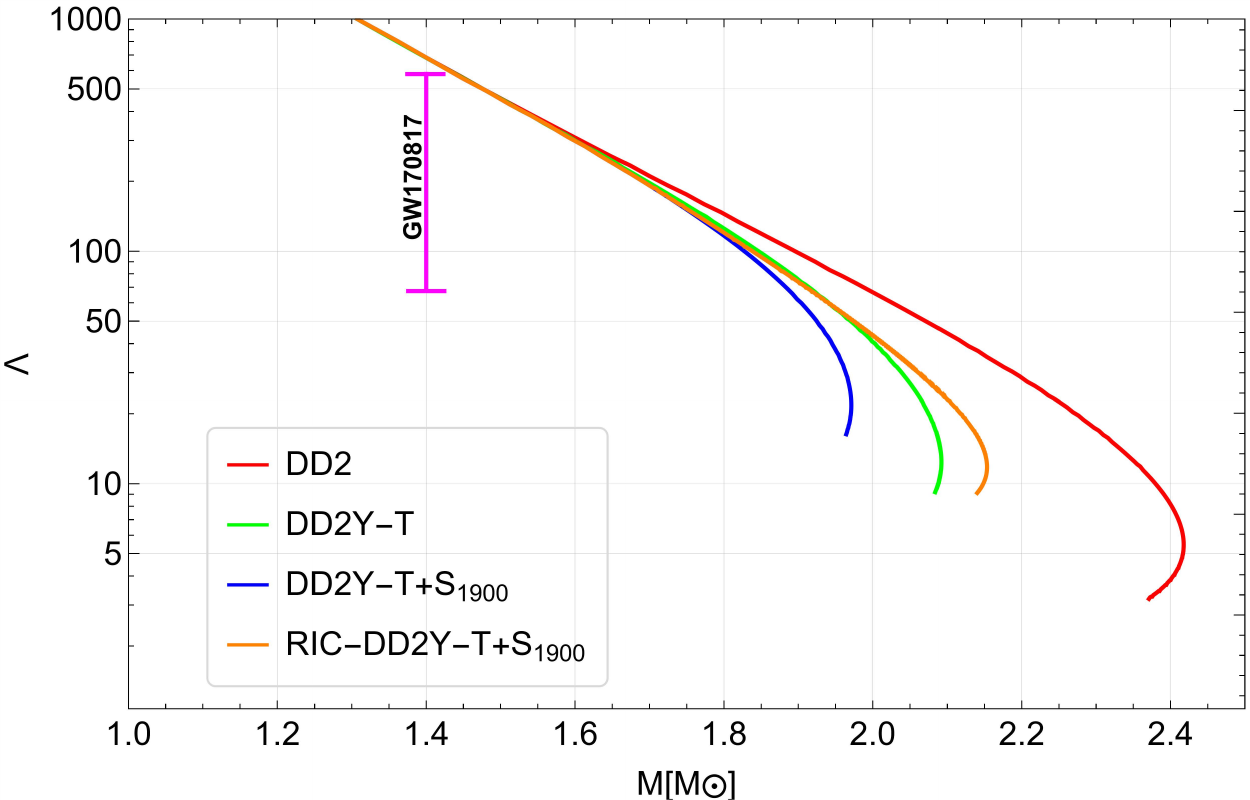}
    \par\medskip
    \centering\textbf{Third scenario}
  \end{minipage}
  \caption{Similar to figure \ref{fig:M-R}, but for the tidal deformability versus the mass of the object. The observational constraint from the LIGO–Virgo collaboration for a $1.4~M_\odot$ NS, derived from the GW170817 event, is indicated by a magenta line.}
  \label{fig:Lambda}
\end{figure*}

 Multi-messenger observations of NSs provide a significant opportunity to probe high-density matter in compact objects. In this regard, GW detectors, LIGO, Virgo and KAGRA and the Neutron Star Interior Composition ExploreR (NICER) telescope play a noteworthy role. GW observations of NS mergers provide a constraint on the tidal deformability parameter ($\Lambda$), which denotes the deformation of the object in a binary system. 
The famous GW170817 event \cite{LIGOScientific:2017vwq} introduced a bound on the dimensionless tidal deformability of $ 1.4 M_\odot$ NS, $70\leq\Lambda_{1.4}\leq580$ \cite{LIGOScientific:2018cki}, which is widely applied to constrain the EOS of high-density matter in NSs. Moreover, considering X-ray pulse profile modeling for pulsars, the NICER telescope has measured the mass and radius of NSs at a high precision level. The updated and recently obtained mass-radius parameter of spaces for  pulsars, PSR
J0030+0451 \cite{Vinciguerra:2023qxq},  PSR J0740+6620 \cite{Salmi:2024aum}, PSR J0437-4715 \cite{Choudhury:2024xbk} along with the latest one PSR J0614-3329 \cite{mauviard2025nicerview14solarmass} provided notable mass and especially radius limits for NSs. In this section, considering the above-mentioned astrophysical limits, we are going to examine different potential structures in NSs, including nucleons, hyperons, DM and phase transition to QM. Regarding the three scenarios applied in this study, figure \ref{fig:M-R} and figure \ref{fig:Lambda} indicate the mass-radius profiles and the tidal deformability variation in terms of mass, respectively.

In the left panel of figure \ref {fig:M-R}, mass-radius curves for the first scenario (RIC-DD2Y-T+S) are shown, for which different masses of S particles are taken into account as can be seen in the legend. It is illustrated that lower masses of DM particles give smaller radii around $1.4M_\odot$, which would be a notable point regarding $f$-mode analysis. Moreover, it is seen that the last two NICER results (PSR J0437-4715 and  PSR J0614-3329) are more in favor of low mass S particles, while the $2M_\odot$ limit is consistent for all cases considered in this plot. Furthermore, it is worth mentioning that for this scenario, the onset of the phase transition is shifted to higher densities by increasing the mass of DM particles. Looking at the middle panel of figure \ref{fig:M-R}, it is seen that the presence of DM particles inside NS (RIC-DD2Y-T+S) changes the radius significantly in comparison to the case where the S particles are absent in the NS (RIC-DD2Y-T). This feature is owing to the fact that the production of S particles with 
$x=0.03$ considerably alters the onset of the phase transition to lower densities. This is an important point regarding the effect of DM inside NSs and the consistency of the hybrid model with observational constraints, which can lead to noticeable variation in the $f$-mode oscillation frequency. Finally, in the right panel, the effect of introducing new degrees of freedom is examined based on the third scenario. It is seen that the main impact of the presence of the S particles, along with hyperons and deconfined QM, is on the maximum mass of the object. In fact, with the coupling constant set to $x=0.08$, the hadronic EOS becomes stiffer, pushing both the onset of the S component and the deconfinement transition to high densities near the maximum-mass configuration, so that the radii of lower-mass stars remain essentially unchanged. In addition, the phase transition to QM in RIC-DD2Y-T+S increases the maximum mass of the compact object compared to DD2Y-T and DD2Y-T+S, whose EOSs are softened by the inclusion of additional particles, in contrast to DD2, which is the stiffest and contains only nucleons.

\begin{figure*}[htbp]
  \centering
  \begin{minipage}[b]{0.5\linewidth}
    \centering
    \includegraphics[width=\linewidth]{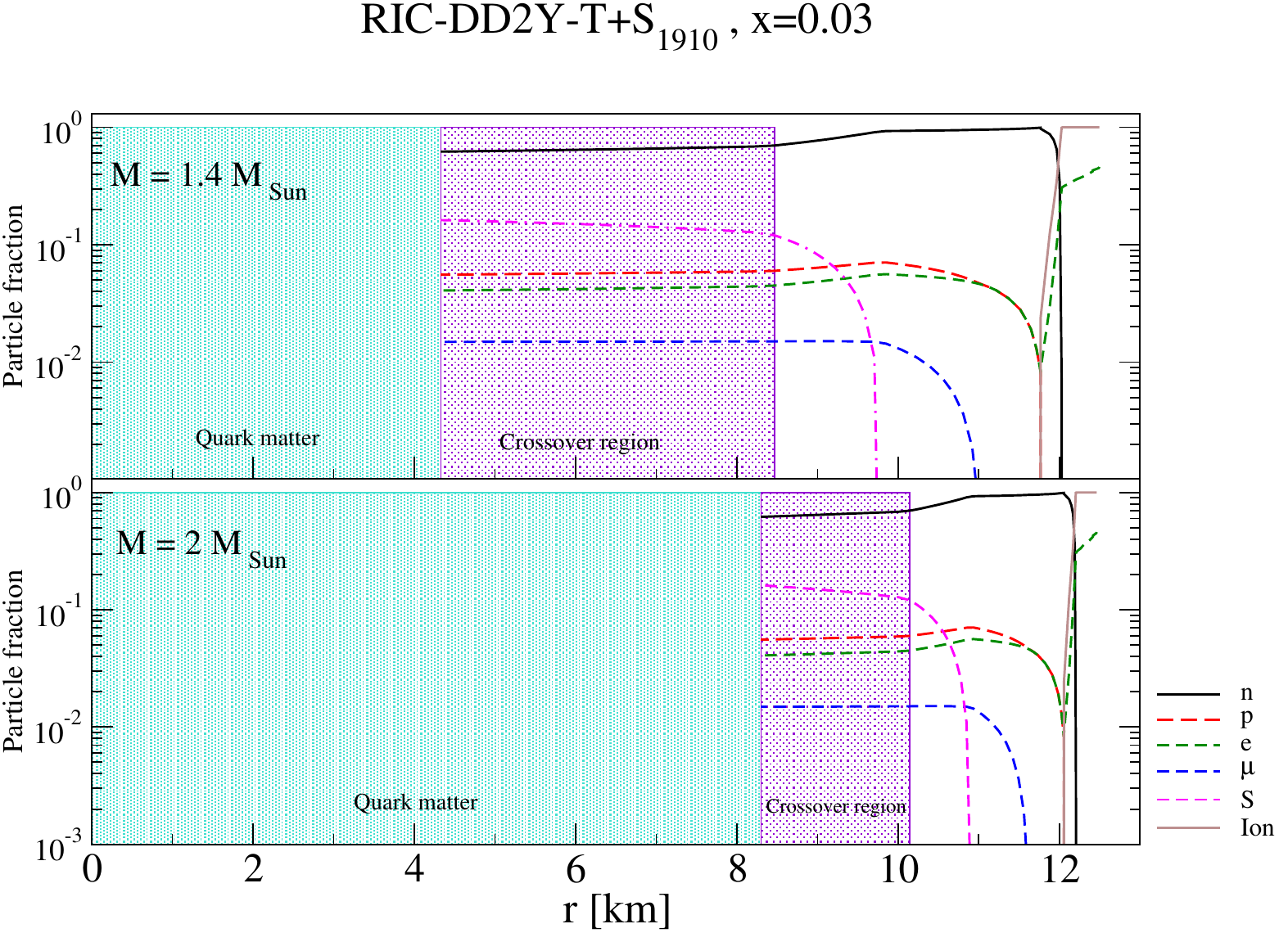}
    \par\medskip
  \end{minipage}\hfill
  \begin{minipage}[b]{0.5\linewidth}
    \centering
    \includegraphics[width=\linewidth]{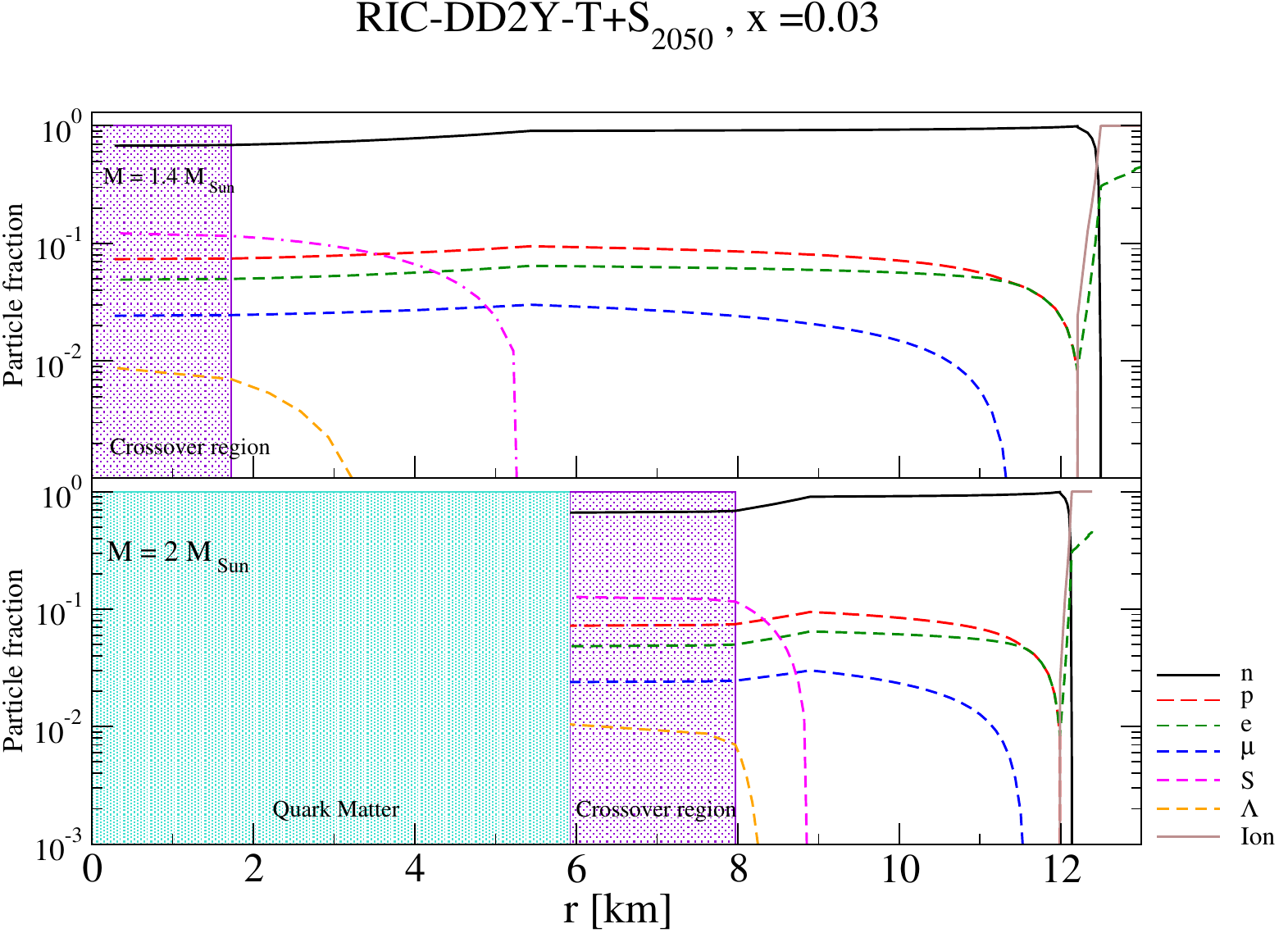}
    \par\medskip
  \end{minipage}\hfill

  \caption{Particle fractions as a function of radial distance from the center of the star for the first scenario. {The effect of different S-particle masses is considered. The left (right) panel corresponds to $m_s=1910$ MeV ($m_s=2050$ MeV), with $x=0.03$ fixed. In both panels, the upper (lower) plot corresponds to a $1.4\,M_{\odot}$ ($2\,M_{\odot}$) NS. The purple and cyan shaded regions indicate the crossover (mixed) phase and the pure quark matter phase, respectively.}}
  \label{fig:profile1}
\end{figure*}

Figure \ref{fig:Lambda} shows the variation of tidal deformability in terms of mass of the object, where the GW170817 constraint is shown by the vertical magenta line. In the left panel, it is seen that for all considered DM particle (S) masses, the $\Lambda_{1.4}$ limit is satisfied well. Furthermore, it should be noted that in the middle panel, where the effect of S production is examined, DM particles have a modifying impact on the hybrid model. This leads to an EOS for NSs which includes all potential exotic degrees of freedom from hyperon, DM to quark matter, while being consistent with the astrophysical constraint. Finally, as illustrated in the right panel, given our modeling for the third scenario, adding new degrees of freedom does not affect the tidal deformability parameter, since the radius for a $1.4M_\odot$ NS remains unchanged.


In conclusion, in this section, we analyze three different scenarios based on the latest multi-messenger observational limits for NSs and check the impacts of having new degrees of freedom, particularly DM production. It is shown that considering S particles as bosonic DM particles with $x=0.03$ (first scenario), makes the hybrid model consistent with all available observational constraints. Thus, given this result, in the following we will analyze the oscillation properties of the $f$-mode and the corresponding universal relations.

\subsection{Profile of the stars}

In figures \ref{fig:profile1}, \ref{fig:profile2}, and \ref{fig:profile3}, we present the radial composition profiles of NSs computed for each of the considered scenarios. These profiles illustrate the variation of particle fractions as a function of radial distance from the stellar center, covering the entire internal structure, from the dense inner core to the outermost regions of the crust. In the core, where densities are highest, a rich mixture of baryonic and, potentially, exotic components such as hyperons, S DM or deconfined QM can be present, depending on the EOS and model parameters. Moving outward, the composition gradually evolves, with heavier particles vanishing as the density decreases. At the crustal layers, the matter becomes increasingly neutron-rich, ultimately transitioning to a region composed primarily of electrons and fully ionized nuclei. These detailed profiles provide crucial insight into how different microscopic compositions manifest within the macroscopic structure of NSs.

It is worth noting that the profiles in each panel are shown for both $1.4M_\odot$ and $2M_\odot$ stars, corresponding to the upper and lower panels, respectively. As expected, the lower-mass star has a larger radius and is therefore less compact. figure \ref{fig:profile1} presents the internal profiles for the first scenario introduced in the previous section. The left panel corresponds to $m_S = 1910$ MeV, {representing one of the lower (and therefore more probable)} masses considered in this scenario, while the right panel shows results for $m_S = 2050$ MeV, the highest value examined {for pedagogical purposes,i.e., to illustrate how a high-mass bosonic DM candidate with double strangeness influences the radial composition profile of NS.}

For the lighter S {($m_S = 1910$ MeV)}, both the onset of S DM and the deconfinement occur at larger radii and hence at lower densities, which prevents the appearance of hyperons. In contrast, for the heavier S {($m_S = 2050$ MeV)}, deconfinement begins at smaller radii and higher densities, allowing $\Lambda$ hyperons to emerge. As a result, the hadronic phase in this case includes ordinary nuclear matter, hypernuclear matter, and DM.

Moreover, in the case of the $1.4~M_\odot$ star shown in the right panel, the central density is insufficient to reach the pure deconfined QM phase. However, the transition to QM begins within the crossover region. This intermediate region, often referred to as the mixed phase, is assumed to contain a combination of hadronic matter and QM. {However, it should be noted that the precise structure of the mixed phase, labeled as the "crossover region" in the figures, is not well determined, and no definitive statements can be made about its detailed composition. For this reason, we have extended the particle fraction lines into this region to indicate the maximum possible presence of each species within the star. Once the pure QM phase starts, the hadronic particle fraction lines no longer appear.}

\begin{figure*}[htbp]
  \centering
  \begin{minipage}[b]{0.5\linewidth}
    \centering
    \includegraphics[width=\linewidth]{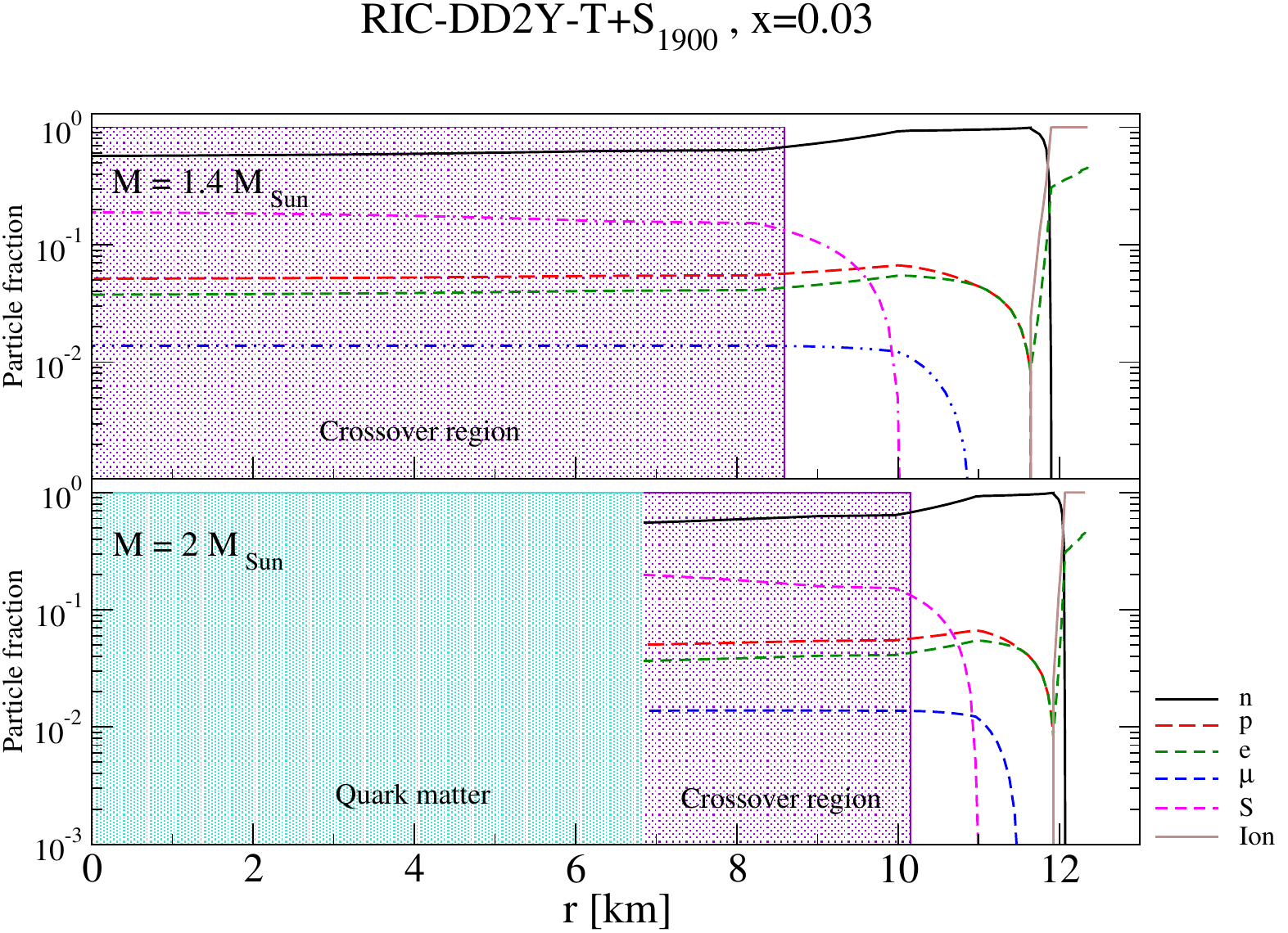}
    \par\medskip
  \end{minipage}\hfill
  \begin{minipage}[b]{0.5\linewidth}
    \centering
    \includegraphics[width=\linewidth]{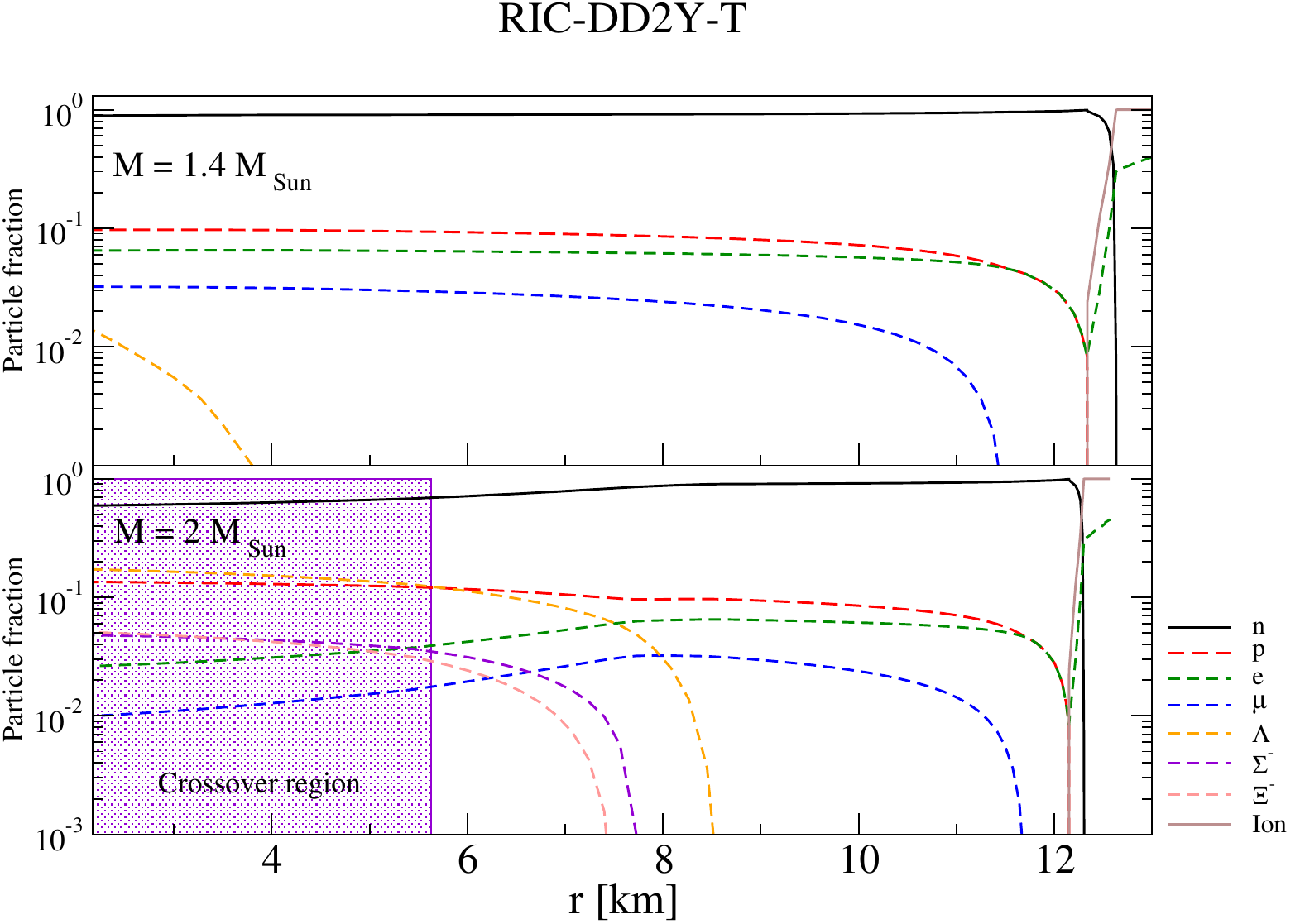}
    \par\medskip
  \end{minipage}\hfill

  \caption{{Similar to Fig.~\ref{fig:profile1}, but for the second scenario in which the impact of 
S-particle production is examined. The left panel corresponds to the model that includes 
S particles ($m_s=1900$ MeV), while the right panel shows the corresponding case without DM (S) in the NS.}}
  \label{fig:profile2}
\end{figure*}

Figure \ref{fig:profile2} presents the internal structure profiles for the second scenario introduced in the previous section. This scenario is designed to investigate the impact of the emergence of S DM within the hadronic phase on the hybrid models. The left panel displays the profile corresponding to the hybrid EOS {incorporating S DM in the hadronic sector (RIC-DD2Y-T+S$_{1900}$),} with a S mass of 
$m_S=1900$ MeV and $x=0.03$. {It is worth mentioning that the profiles have been mainly plotted for a S mass around $1900~\text{MeV}$ to remain consistent with the favored mass range obtained in our previous study \cite{Shahrbaf:2024gdm}.} The right panel shows the profile for the hybrid EOS with hyperons but without the inclusion of S DM (RIC-DD2Y-T).
As shown in the left panel, the upper plot corresponds to the $1.4M_\odot$ star. Since this star is less compact, the central density does not reach the threshold for forming a pure deconfined quark matter core. In contrast, for the more massive $2M_\odot$ star shown in the lower panel, the internal structure displays a sequential layering: starting from the surface toward core, we observe pure nucleonic matter, followed by a region containing S DM, then the crossover region where hadronic and quark phases coexist, and finally the pure quark matter core. Due to the early onset of light S DM in this configuration, the appearance of $\Lambda$ hyperons is suppressed.

The right panel offers an interesting contrast. In the absence of S DM, hyperons are no longer suppressed by competition with the DM degrees of freedom. As a result, in the $1.4M_\odot$ star (upper panel), $\Lambda$ hyperons are present, and in the $2M_\odot$ star (lower panel), a richer spectrum of hyperonic degrees of freedom emerges, including $\Lambda$, $\Sigma^-$, and $\Xi^-$. This behavior is attributed both to the absence of S DM and to the increased compactness of the higher-mass star, which allows the necessary conditions for hyperon formation to be met more easily.

Figure \ref{fig:profile3} shows the particle profiles for the third scenario. This figure allows us to compare how different internal structures develop depending on the EOS. The lower right panel displays a purely nucleonic star (DD2), while the lower left panel shows a hyperonic star (DD2Y-T). The upper right panel corresponds to a pure hadronic star that includes S DM (DD2Y-T+$\text{S}_{1900}$, $x = 0.08$), and the upper left panel presents a hybrid star with all possible components, i.e., nucleons, hyperons, S DM, and deconfined QM (RIC-DD2Y-T+$\text{S}_{1900}$, $x = 0.08$).
This setup allows us to clearly observe how the appearance of new degrees of freedom, such as hyperons, S DM, or QM, affects the internal composition of the star. It also illustrates under what density conditions each type of particle can appear within the star, based on their radial distance from the center.

 In the DD2 model, the star's core, at the highest densities, contains only nucleons, electrons, and muons. In the lower-density crust region, as expected, the composition includes ions and electrons. For DD2Y-T, hyperons also appear in the core. By comparing the lower and upper panels, we see that in the $1.4M_{\odot}$ star, only the $\Lambda$ hyperon emerges. In contrast, for the $2M_{\odot}$ star, as the density increases (i.e., as we move inward toward the center), the $\Lambda$, $\Sigma^-$, and $\Xi^-$ hyperons appear sequentially. This reflects the higher compactness of the $2M_{\odot}$ star, where denser conditions allow for the emergence of heavier particles.

In the DD2Y-T+$\text{S}_{1900}, x=0.08$ case, the base EOS (DD2Y-T) is modified by including S DM with a mass of $m_S = 1900$ MeV and a slope parameter $x = 0.08$. {Although this corresponds to one of the lightest S-mass values,} the relatively large coupling constant $x$ increases the stiffness of the EOS and shifts the onset of S production to higher densities.
 As a result, in the $1.4M_{\odot}$ star, the S DM does not appear, and among strange particles, only the $\Lambda$ hyperon is present. However, in the $2M_{\odot}$ star, the central density becomes high enough for the appearance of $\Lambda$, S, $\Sigma^-$, and $\Xi^-$ in that order. Notably, compared to the scenario with $x=0.03$, the higher repulsion introduced by $x=0.08$ shifts the S onset to higher densities, reversing the order of appearance between S and $\Lambda$.

Finally, in the RIC-DD2Y-T+$\text{S}_{1900}$, $x=0.08$ case, the EOS remains stiff, which delays the onset of quark deconfinement to higher densities. For the $1.4M_{\odot}$ star, only the $\Lambda$ hyperon appears, as in DD2Y-T+$\text{S}_{1900}$, and the high value of $x$ prevents the emergence of S DM. In the $2M{\odot}$ star, all strange components, $\Lambda$, S, $\Sigma^-$, and $\Xi^-$, are present, but they compete with QM. Consequently, their particle fractions remain lower than in DD2Y-T+$\text{S}_{1900}$ due to the onset of the QM transition. Since the deconfinement threshold lies at very high densities, even in the $2M_{\odot}$ star, only the crossover region appears, and the pure QM phase is not reached. This behavior mirrors that of the RIC-DD2Y-T model in the second scenario, where the stiff hadronic EOS causes a delayed phase transition.

An important feature observed in the profile figures is that, whenever $\Sigma^-$ and $\Xi^-$ hyperons begin to appear, deleptonization sets in, resulting in a noticeable decrease in the electron and muon fractions within the NS. 

Another key conclusion of this section is the competition between hyperons and S DM, which is a doubly-strange boson. In our scenarios, for S masses around $1900~\text{MeV}$ and $x=0.03$, the preferred values based on our previous work \cite{Shahrbaf:2024gdm}, hyperons are largely replaced by S DM. However, for higher values of the S mass or larger values of the slope parameter $x$, the onset of S DM shifts to higher densities, allowing hyperons to emerge earlier and dominate the composition, leading to the formation of a hyperonic star with DM.

\begin{figure*}[htbp]
  \centering
  \begin{minipage}[b]{0.5\linewidth}
    \centering
    \includegraphics[width=\linewidth]{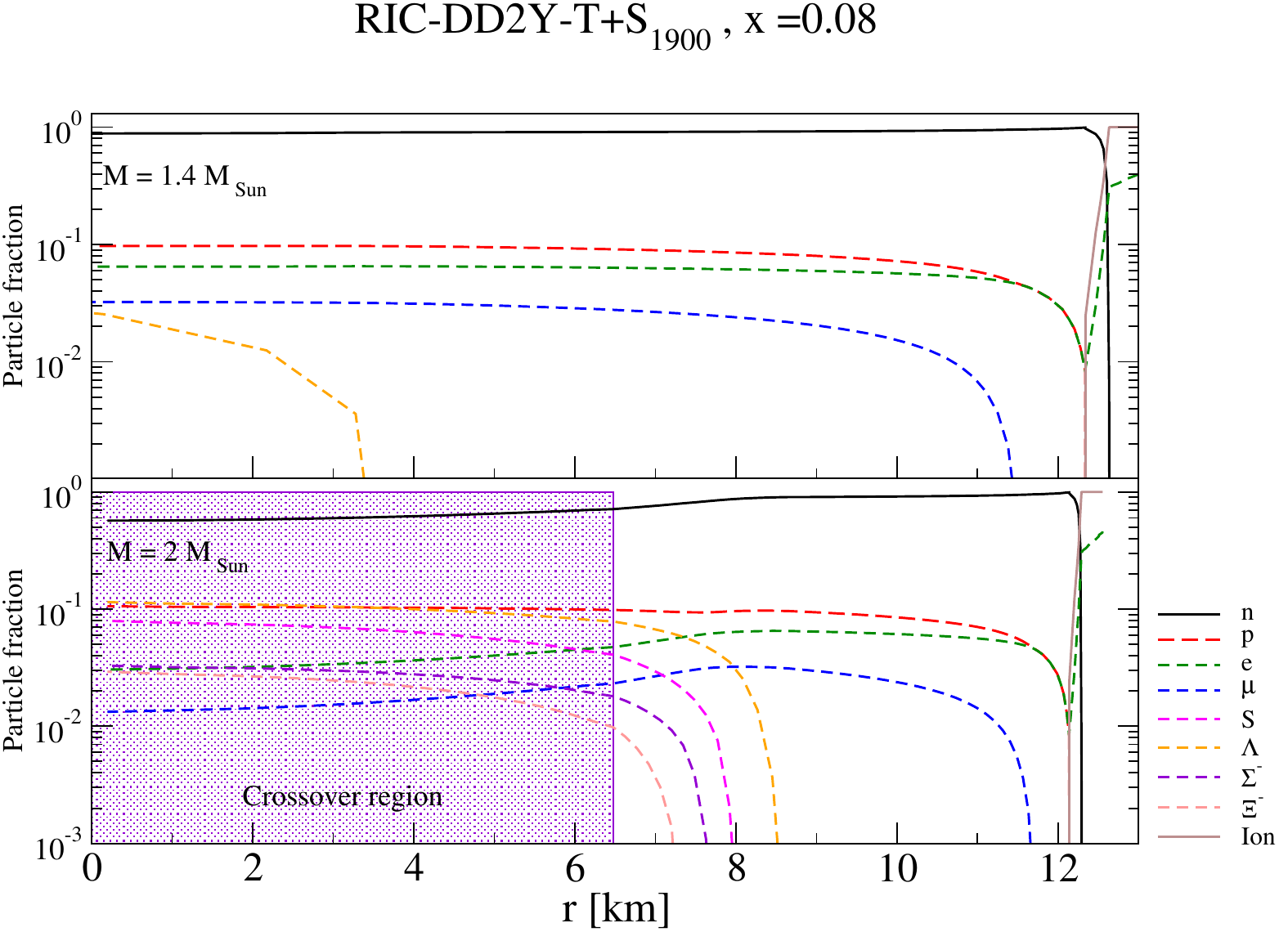}
    \par\medskip
  \end{minipage}\hfill
  \begin{minipage}[b]{0.5\linewidth}
    \centering
    \includegraphics[width=\linewidth]{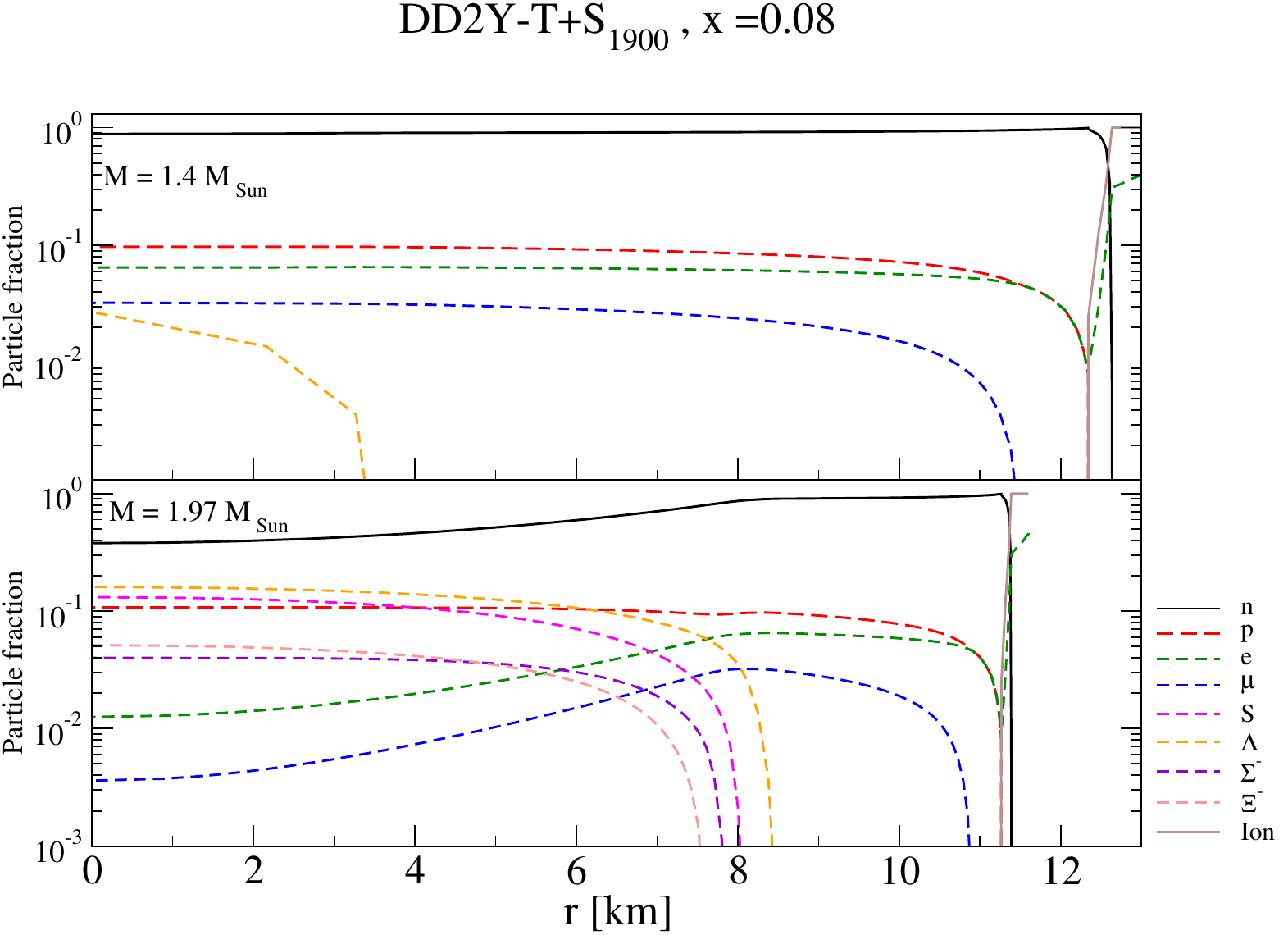}
    \par\medskip
     \end{minipage}\hfill
    \\
    \begin{minipage}[b]{0.5\linewidth}
    \centering
    \includegraphics[width=\linewidth]{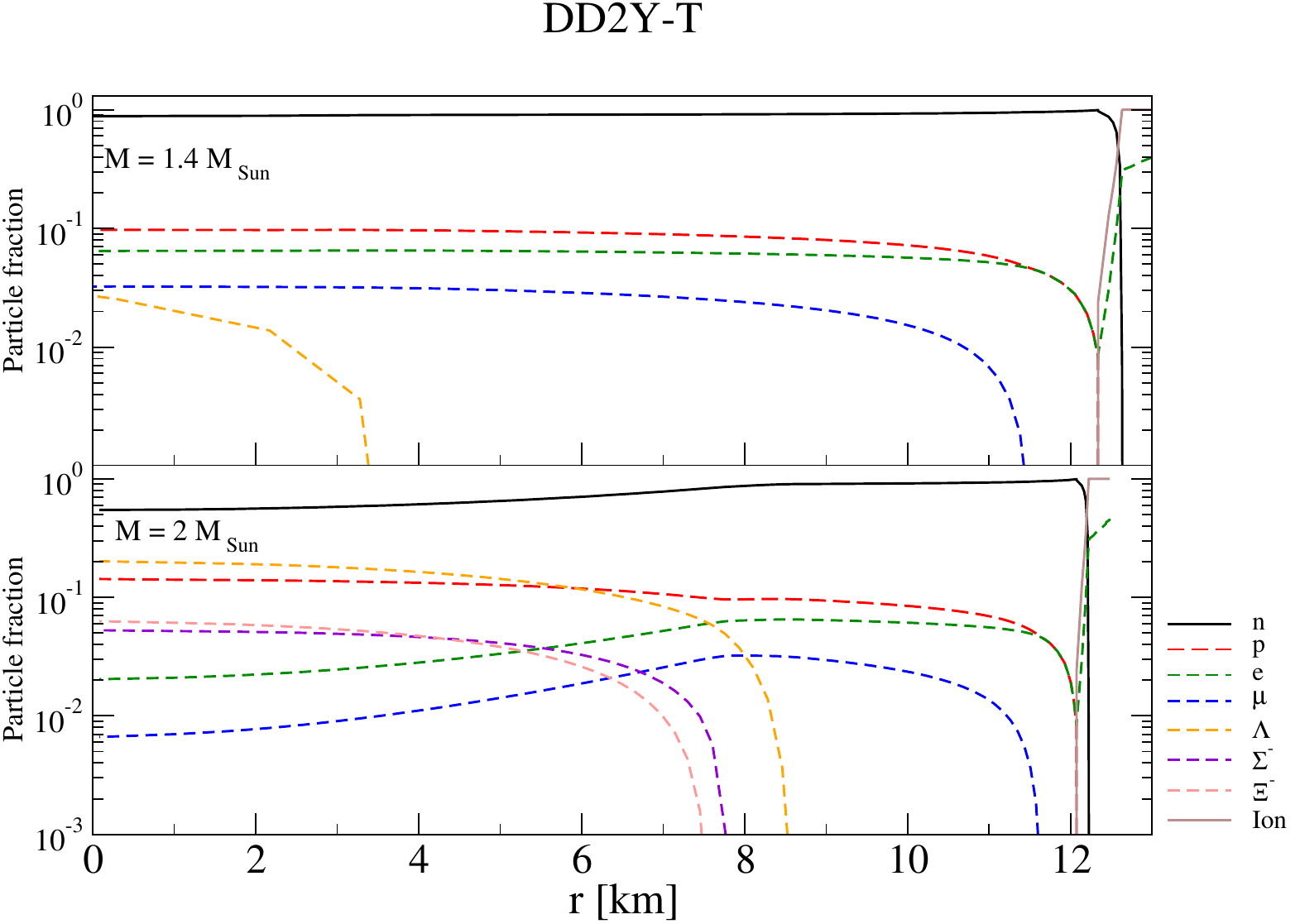}
    \par\medskip
  \end{minipage}\hfill
  \begin{minipage}[b]{0.5\linewidth}
    \centering
    \includegraphics[width=\linewidth]{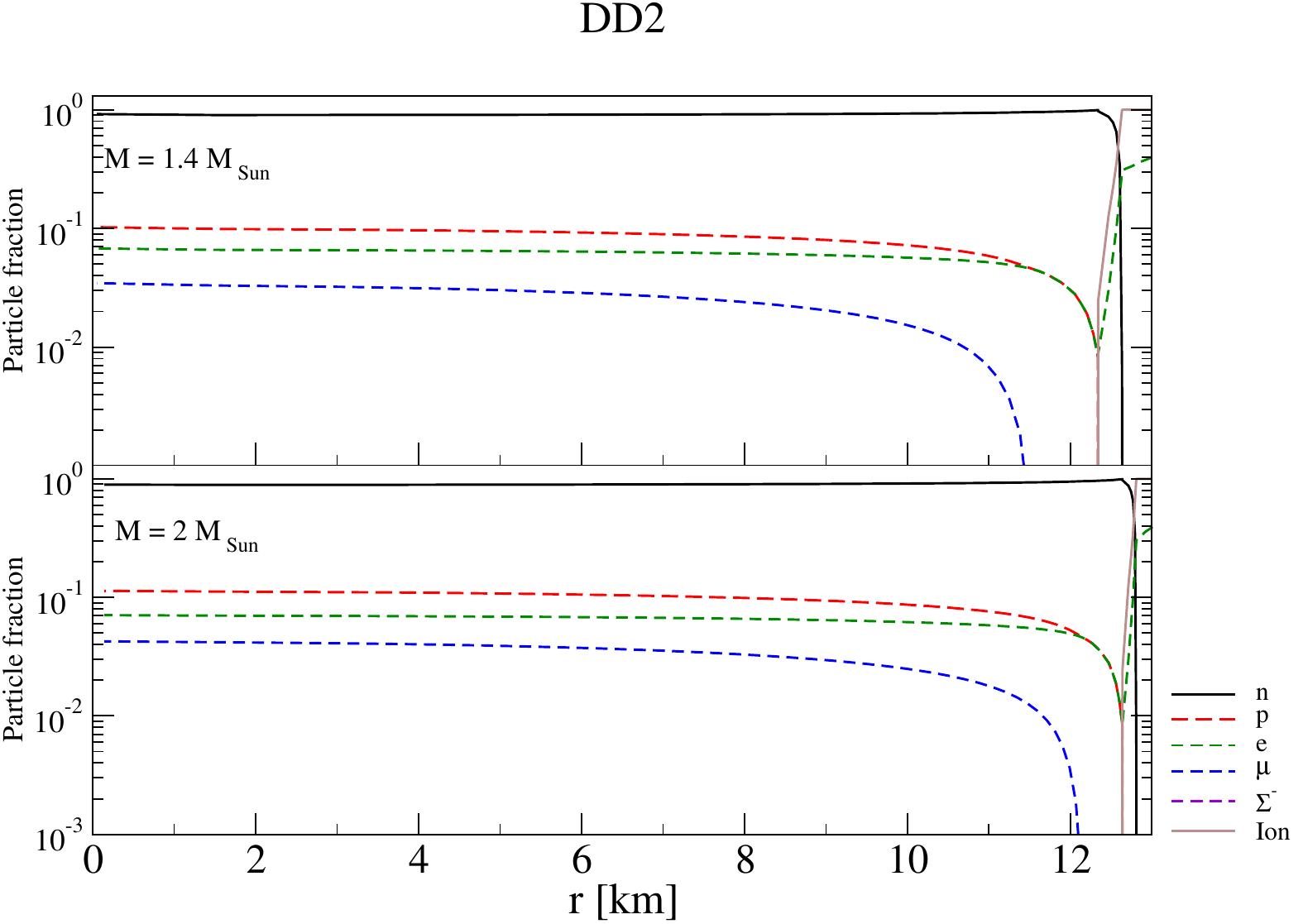}
    \par\medskip
  \end{minipage}\hfill

  \caption{{Similar to Fig.~\ref{fig:profile1}, but for the third scenario, where the impact of different degrees of freedom in the NS interior is investigated. The lower-right panel shows the purely nucleonic model (DD2), while the lower-left panel corresponds to the hyperonic case (DD2Y-T). In the upper row, the right panel corresponds to the model in which the S is produced in the hyperonic NS (DD2Y-T$+S_{1900}$), whereas the left panel shows the corresponding hybrid model (RIC-DD2Y-T$+S_{1900}$), where quark matter can appear in the NS core.}}
  \label{fig:profile3}
\end{figure*}

\subsection{$f$-mode frequency}

\begin{figure*}[htbp]
  \centering
  \begin{minipage}[b]{0.32\linewidth}
    \centering
    \includegraphics[width=\linewidth]{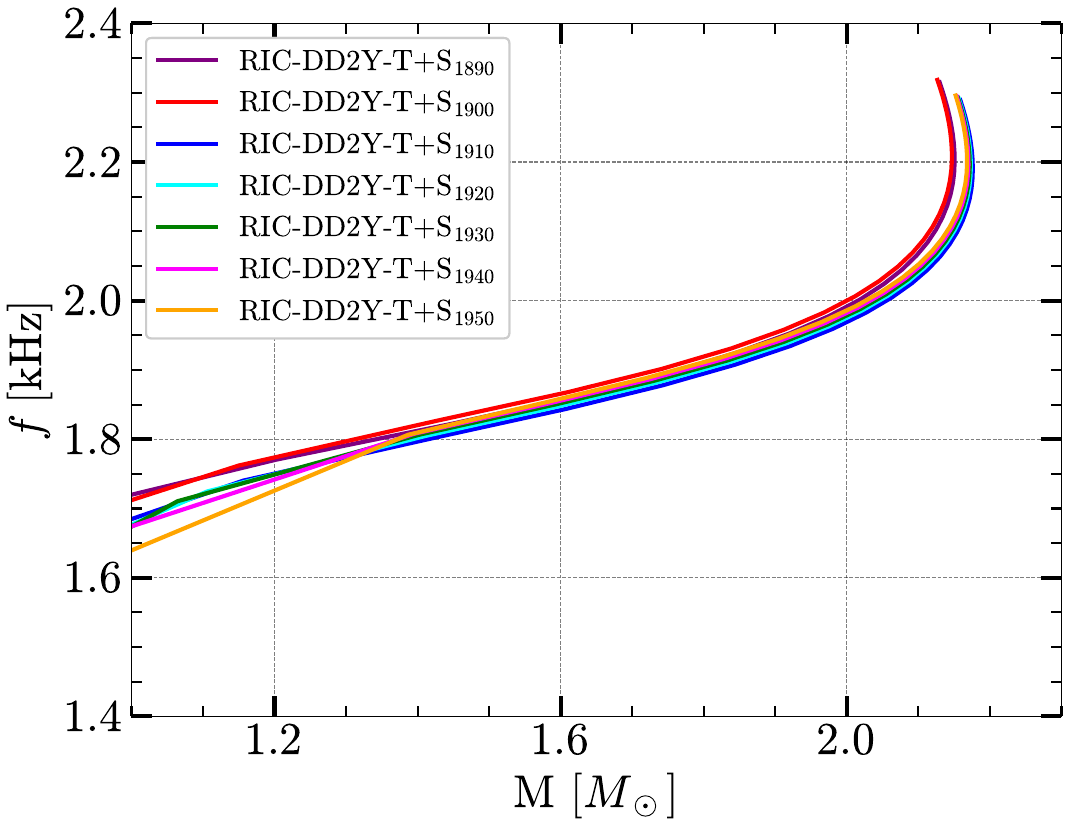}
    \par\medskip
    \centering\textbf{(a)}
  \end{minipage}\hfill
  \begin{minipage}[b]{0.32\linewidth}
    \centering
    \includegraphics[width=\linewidth]{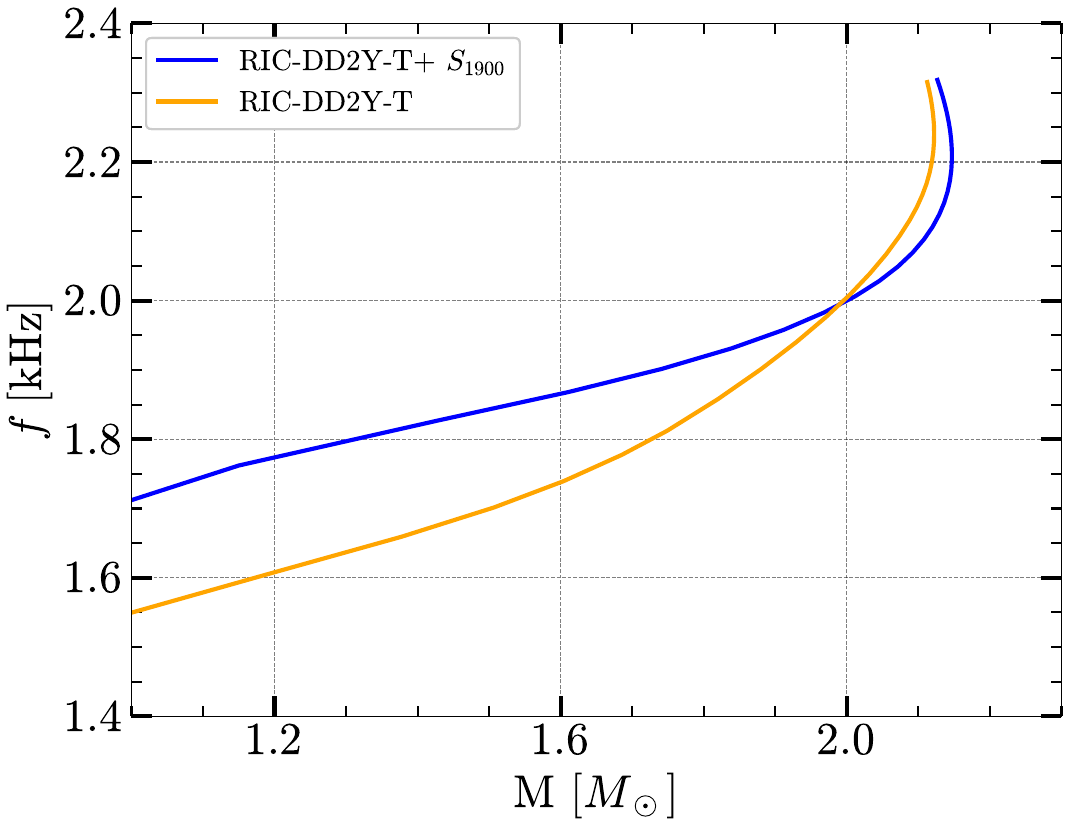}
    \par\medskip
    \centering\textbf{(b)}
  \end{minipage}\hfill
  \begin{minipage}[b]{0.32\linewidth}
    \centering
    \includegraphics[width=\linewidth]{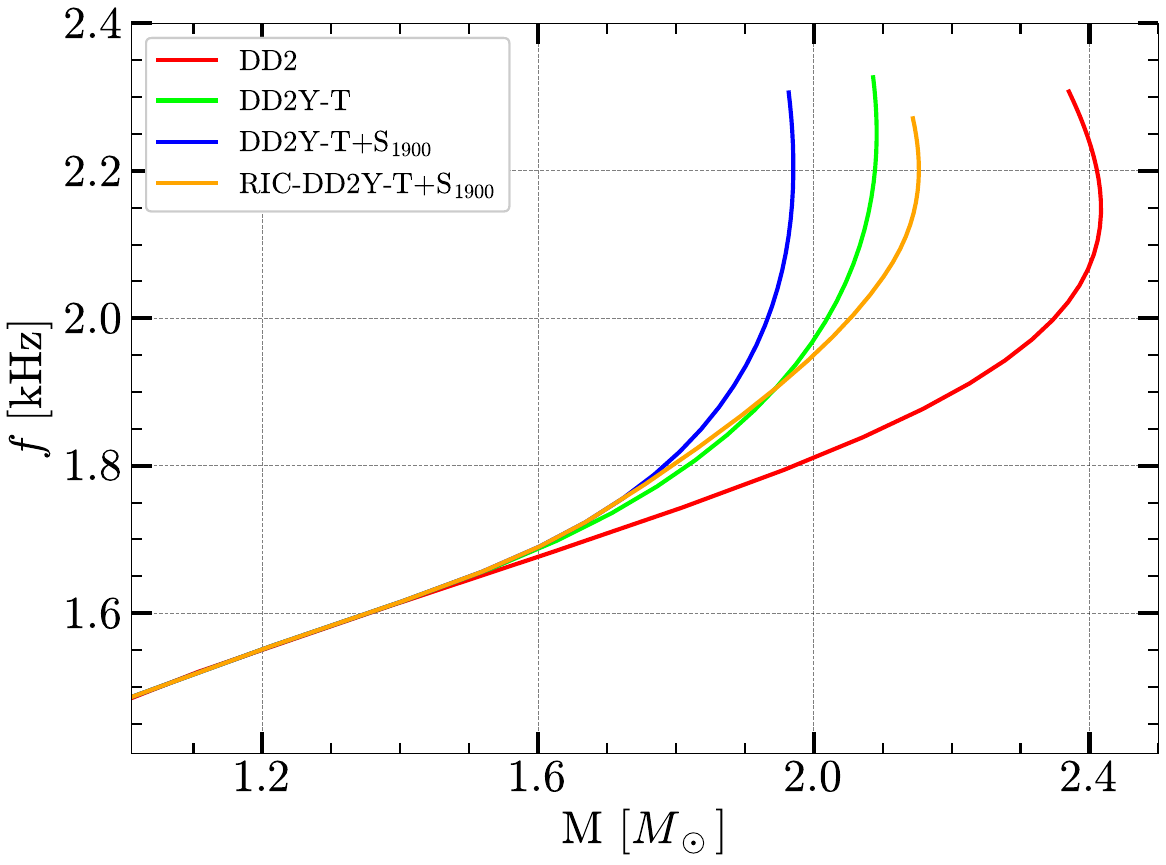}
    \par\medskip
    \centering\textbf{(c)}
  \end{minipage}
  \caption{(a) Frequency in terms of the mass for DM admixed hybrid stars with various S masses. (b) Frequency in terms of the mass of the object for hybrid stars without DM (orange line) and with DM for $m_s=1900$ MeV and $x=0.03$ (blue line). (c) Frequency in terms of the mass for the DM admixed {hybrid star (orange line), along with nucleonic star (red line), hyperonic star (green line), and DM admixed hyperonic star (blue line) for $m_s=1900$ MeV and $x=0.08$.}}
  \label{fig:frequency_mass_combined}
\end{figure*}

We now turn to the analysis of the $f$-mode oscillations, focusing on how the f-mode oscillations vary with the gravitational mass and tidal deformability of NSs, as well as the associated damping time. The variation of the $f$-mode frequency with the gravitational mass of NSs is displayed in figure~\ref{fig:frequency_mass_combined}.

The left panel of this figure displays the \( f \)-mode frequency in relation to NS mass (\(M_{\odot}\)) for the hybrid EOSs RIC-DD2Y-T+S, which incorporate a smooth crossover phase transition to deconfined QM within a hadronic envelope composed of nucleons and hyperons, further admixed with bosonic DM (S particles). The colored lines represent different S-particle masses ranging from 1890 to 1950 MeV, with a fixed coupling strength \( x = 0.03 \). At lower masses (below \(1.4\,M_\odot\)), a clear spread among the models emerges. Heavier DM particles, such as \(m_S=1950\) MeV (orange line), tend to yield the lowest oscillation frequencies, while lighter DM particles like \(m_S=1890\) MeV and \(m_S=1900\) MeV exhibit the highest frequencies. {At \(1.4\,M_\odot\), all models converge to similar frequencies. Beyond this canonical mass, the \(m_S=1910\) MeV model shows the lowest value (1.797 kHz), reflecting a stiffer EOS that supports the highest maximum mass (\(2.176\,M_\odot\)), whereas the \(m_S=1900\) MeV model reaches the highest frequency (1.820 kHz), consistent with a softer EOS and the lowest maximum mass (\(2.147\,M_\odot\)), indicative of a more compact configuration. These differences are also quantitatively supported by Table~\ref{tab:GR_frequencies_damping}.

\begin{table}[htbp]
\centering
\caption{The $f$-mode frequencies ($f$) in kHz and damping times ($\tau$) in seconds for the first scenario, evaluated at 1.4, 1.8, and the maximum mass (in $M_{\odot}$).}
\label{tab:GR_frequencies_damping}
\begin{tabular}{cccc}
\hline
Model & Mass ($M_\odot$) & $f$ (kHz) & $\tau$ (s) \\
\hline
\multirow{3}{*}{RIC-DD2Y-T+$S_{1890}$}
  & 1.4   & 1.812 & 0.2154 \\
  & 1.8   & 1.911 & 0.1608 \\
  & 2.151 & 2.200 & 0.1370 \\
\hline
\multirow{3}{*}{RIC-DD2Y-T+$S_{1900}$}
  & 1.4   & 1.820 & 0.2130 \\
  & 1.8   & 1.919 & 0.1596 \\
  & 2.147 & 2.203 & 0.1367 \\
\hline
\multirow{3}{*}{RIC-DD2Y-T+$S_{1910}$}
  & 1.4   & 1.797 & 0.2179 \\
  & 1.8   & 1.895 & 0.1627 \\
  & 2.176 & 2.190 & 0.1383 \\
\hline
\multirow{3}{*}{RIC-DD2Y-T+$S_{1920}$}
  & 1.4   & 1.802 & 0.2159 \\
  & 1.8   & 1.901 & 0.1618 \\
  & 2.173 & 2.193 & 0.1382 \\
\hline
\multirow{3}{*}{RIC-DD2Y-T+$S_{1930}$}
  & 1.4   & 1.806 & 0.2143 \\
  & 1.8   & 1.904 & 0.1613 \\
  & 2.172 & 2.194 & 0.1381 \\
\hline
\multirow{3}{*}{RIC-DD2Y-T+$S_{1940}$}
  & 1.4   & 1.809 & 0.2133 \\
  & 1.8   & 1.908 & 0.1608 \\
  & 2.170 & 2.196 & 0.1380 \\
\hline
\multirow{3}{*}{RIC-DD2Y-T+$S_{1950}$}
  & 1.4   & 1.810 & 0.2133 \\
  & 1.8   & 1.911 & 0.1603 \\
  & 2.168 & 2.197 & 0.1378 \\
\hline
\end{tabular}
\end{table}

For a better understanding of the behavior of frequencies, one should also consider both figure~\ref{fig:p-e} and figure~\ref{fig:M-R}. In figure~\ref{fig:p-e} it is shown that the stiffness of the EOSs changes in three distinct regions. At low energy densities (\(\epsilon < 300\) $MeV fm^{-3}$), the hybrid EOS is governed by the hadronic part affected by the mass of \(S\) DM; therefore, the lighter the DM mass, the softer the EOS. In the intermediate densities (\(300 \lesssim \epsilon \lesssim 500\) $MeV fm^{-3}$), the parameterization of the mixed-phase region (RIC) becomes dominant, and the ordering of the stiffness among the models changes. This pronounced change in the stiffness and, therefore, in the sequence of $f$-mode frequencies, particularly in the intermediate-density region, reflects the non-monotonic and complex influence of the DM particle mass and the RIC parameterization on the hadron-quark crossover.

 At high energy densities (\(\epsilon \gtrsim 500\) $MeV fm^{-3}$) in figure~\ref{fig:p-e}, all curves converge, since at such densities the QM EOS governs the hybrid star. It is worth noting that the two softest hadronic models (\(m_S=1890, 1900\) MeV) are combined with one QM EOS (characterized by \(\eta_D=0.70\), \(\eta_V=0.10\), and \(C_s^2=0.44\)), while the remaining hadronic models are matched with the another one, slightly stiffer QM EOS (\(\eta_D=0.70\), \(\eta_V=0.11\), and \(C_s^2=0.44\)). The details of the different parameterizations and the criteria for choosing the EOS are discussed extensively in our previous work \cite{Shahrbaf:2024gdm}. The manifestation of these parameterizations for the hadronic EOS, mixed phase, and QM EOS can also be seen in figure~\ref{fig:M-R}. Around \(1.4\,M_\odot\), all separate M-R curves tend to converge toward the maximum mass; in particular, the two softest lines almost overlap, while the other curves also merge due to their similar QM EOS.}


The middle Panel of figure~\ref{fig:frequency_mass_combined} compares the $f$-mode frequencies of a hybrid EOS RIC-DD2Y-T without DM (orange) with a DM admixed configuration (blue curve) with $m_S = 1900$~MeV and a DM coupling constant $x = 0.03$. At the lower mass end ($M \approx 1.1M_\odot$), the DM admixed model exhibits a higher oscillation frequency ($f \approx 1.70$~kHz) compared to the one without DM EOS($f \approx 1.55$~kHz).
Indeed, at lower masses, corresponding to lower densities, the hadronic EOS plays the dominant role. The EOS, including DM is softer than the one without DM. This behavior implies that a stiffer EOS, which produces a larger and less compact star, naturally leads to a lower characteristic $f$-mode frequency compared to a softer EOS, which results in a smaller and more compact star. 

As the stellar mass increases, both frequencies rise monotonically; however, the one without DM steepens more rapidly. Around $M \approx 2.0M_\odot$, the two curves intersect near $f \approx 2.20$~kHz, after which the hybrid EOS without DM overtakes the DM admixed one. This crossing indicates that, although DM enhances the $f$-mode frequency at lower masses, it does not dominate the behavior at high densities, where the hybrid star without DM can sustain larger $f$-mode oscillations approaching the maximum mass limit. At these high densities, the RIC-DD2Y-T EOS becomes increasingly populated with hyperons, as shown in the right panel of figure~\ref{fig:profile2}. The resulting increase in hyperon fractions softens the EOS, particularly in combination with a delayed deconfinement onset, thereby influencing the oscillation frequencies.
 These trends are further supported by Table~\ref{tab:hybrid_frequency_damping_comparison}, where it is evident that the $f$-mode frequencies at $1.4M_\odot$ and $1.8M_\odot$ are higher for the DM admixed stars but become lower near the maximum mass. 

\begin{table}[htbp]
\centering
\caption{$f$-mode frequencies ($f$) in kHz and damping times ($\tau$) in seconds for the second scenario.}
\begin{tabular}{cccc}
\hline
Model & Mass (M$_\odot$) & $f$ (kHz) & $\tau$ (s) \\
\hline
\multirow{3}{*}{RIC-DD2Y-T+ $S_{1900}$} 
    & 1.4   & 1.82 & 0.213 \\
    & 1.8   & 1.92 & 0.160 \\
    & 2.14 & 2.20 & 0.137 \\
\hline
\multirow{3}{*}{RIC-DD2Y-T} 
    & 1.4   & 1.67 & 0.252 \\
    & 1.8   & 1.85 & 0.169 \\
    & 2.12 & 2.24 & 0.135 \\
\hline
\end{tabular}
\label{tab:hybrid_frequency_damping_comparison}
\end{table}

Right panel of figure~\ref{fig:frequency_mass_combined} shows the $f$-mode frequency for four NS EOS models: hadronic DD2 (red), which includes only nucleons; hadronic DD2Y-T (green), which includes nucleons and hyperons; hadronic DD2Y-T+S$_{1900}$ (blue), which additionally incorporates bosonic S DM with a mass of 1900~MeV and a coupling constant of $x = 0.08$; and RIC-DD2Y-T+S$_{1900}$ (orange), a hybrid EOS model that features a phase transition to a quark core surrounded by hadronic matter composed of nucleons, hyperons, and bosonic DM, with mass of 1900 MeV and a coupling constant of $x = 0.08$. It is worth noting that in this scenario, for the hadronic EOS including DM, which represents the softest EOS, the coupling constant has been fixed at $x=0.08$, rather than $x=0.03$ used in previous scenarios, in order to remain consistent with the lower bound on the maximum NS mass from PSR J0740+6620 \cite{Miller:2021qha}.

At 1.4$M_{\odot}$, all models have nearly same frequency which can be also seen in table~\ref{tab:frequency_damping_comparison}. This behavior, also visible in the right panel of figure~\ref{fig:M-R}, indicates that all EOSs in this scenario originate from the same baseline EOS (DD2) and progressively deviate from it by sequentially including additional degrees of freedom: first hyperons, then DM, and finally deconfined QM. 
The purely nucleonic DD2 model serves as the stiffest EOS; therefore, it produces the largest and least compact star, which fundamentally results in the lowest oscillation frequency across the entire density range (also see figure~\ref{fig:p-e}). At 1.8 $M_{\odot}$, the inclusion of hyperons in DD2Y-T leads to a softening of the EOS, resulting in more compact NS. Consequently, their characteristic $f$-mode frequencies are higher compared to those of the stiffer DD2 model. The further inclusion of DM in DD2Y-T+S$_{1900}$, together with the large fractions of hyperons and S at high densities, softens the EOS even more significantly, leading to a more compact star that exhibits the highest $f$-mode frequency among the models considered.
Finally, the hybrid star RIC-DD2Y-T+S$_{1900}$ demonstrates the complex interplay of hyperons, DM, and deconfined QM when the phase transition from DD2Y-T+S$_{1900}$ to QM takes place.
 The emergence of the quark core provides additional pressure support compared to the DM-admixed hadronic matter, making the hybrid star less compact than the DD2Y-T+S$_{1900}$ model, which can be seen from table~\ref{tab:frequency_damping_comparison} and consequently its frequency is lower than that of the DD2Y-T+S$_{1900}$ model. 
 
 \begin{table}[htbp]
\centering
\caption{$f$-mode frequencies ($f$) in kHz and damping times ($\tau$) in seconds for the third scenario.}
\begin{tabular}{cccc}
\hline
Model & Mass (M$_\odot$) & $f$ (kHz) & $\tau$ (s) \\
\hline
\multirow{3}{*}{DD2} 
    & 1.4   & 1.614 & 0.274 \\
    & 1.8   & 1.740 & 0.190 \\
    & 2.41 & 2.141 & 0.161 \\
\hline
\multirow{3}{*}{DD2Y-T} 
    & 1.4   & 1.615 & 0.274 \\
    & 1.8   & 1.789 & 0.181 \\
    & 2.09 & 2.242 & 0.137 \\
\hline
\multirow{3}{*}{DD2Y-T+$S_{1900}$} 
    & 1.4   & 1.615 & 0.274 \\
    & 1.8   & 1.814 & 0.177 \\
    & 1.97 & 2.196 & 0.135 \\
\hline
\multirow{3}{*}{RIC-DD2Y-T+$S_{1900}$} 
    & 1.4   & 1.615 & 0.274 \\
    & 1.8   & 1.804 & 0.179 \\
    & 2.15 & 2.199 & 0.140 \\
\hline
\end{tabular}
\label{tab:frequency_damping_comparison}
\end{table}
 
 Near the maximum-mass configurations, the pronounced enhancement of hyperon fractions in the DD2Y-T model softens the EOS relative to the hybrid EOS. The reduced pressure support at these central densities produces smaller radii (i.e., more compact configurations) for a given gravitational mass, which in turn shifts the $f$-mode frequencies upward. At slightly lower central densities, before hyperons dominate and while the hadron-quark crossover and the RIC parameterization control the stiffness, the ordering of model stiffnesses is reversed and the hybrid EOS is relatively softer, producing the opposite trend in frequencies. This behavior is also evident in the right panel of figure~\ref{fig:M-R}.
 Therefore, the hierarchy of frequencies at a fixed mass directly maps to the hierarchy of stellar compactness, providing a clear GW signature of the softening caused by exotic particles.

 \begin{figure}[htbp]
\centering
	\includegraphics[width=0.7\textwidth]{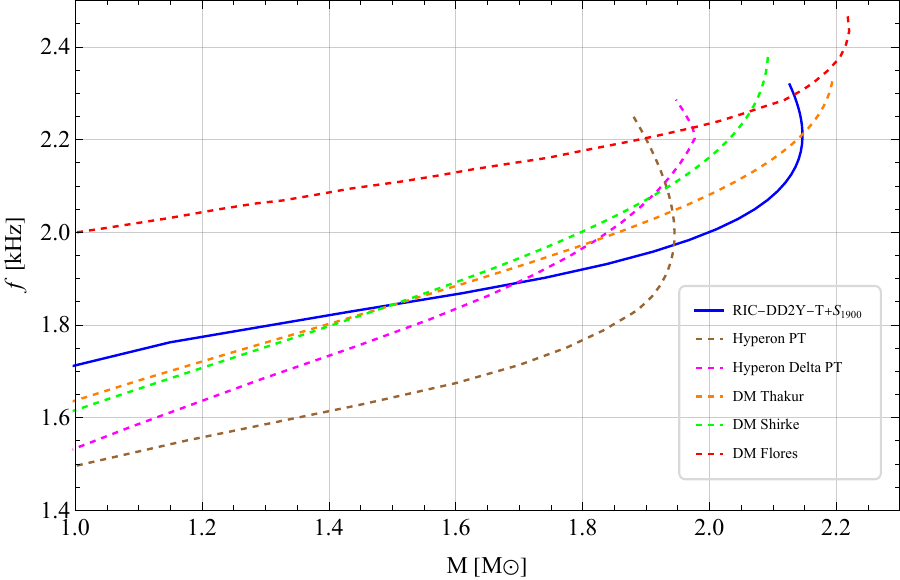}

	\caption{{The f-mode frequency as a function of mass for different EOS models.
The blue solid curve corresponds to our model, i.e. a hybrid NS including hyperons, bosonic S DM
($m_S=1900~\mathrm{MeV}$), and quark matter.
The brown and magenta dashed curves show the hyperonic and $\Delta$-admixed EOSs, respectively, in which a phase
transition (PT) to quark matter occurs at high densities \cite{Rather:2024mtd}.
The orange, green, and red dashed curves correspond to nucleonic EOSs including fermionic DM
\cite{Thakur:2025zhi,Shirke:2024ymc,Flores:2024hts}.}
		\label{fig:f-m compare}
	}
\end{figure}

{For the sake of completeness, in the figure.~\ref{fig:f-m compare}, we compare the $f$-mode frequencies of our hybrid EOS with several other EOS models from the literature. Our model, which includes hyperons, a 1900 MeV S DM component, and deconfined quark matter in addition to ordinary nuclear matter, is shown by the solid blue curve. For comparison, we plot EOSs with hyperons and a phase transition to quark matter (dashed brown and magenta lines) as well as nucleonic EOSs with fermionic DM (dashed orange, green, and red lines). The $f$-mode frequency for each model is computed using the full GR approach, ensuring a consistent comparison. It is seen that the dashed brown line \cite{Rather:2024mtd}, which includes hyperons and a phase transition to quark matter, generally yields the lowest frequencies. When the Delta baryons are added, as shown by the dashed magenta line \cite{Rather:2024mtd}, an additional degree of freedom is introduced, which softens the EOS and increases the $f$-mode frequency. By contrast, when compared with our model, higher frequencies are obtained due to the inclusion of heavy bosonic DM in addition to hyperons and quark matter, resulting in an even softer EOS, particularly before the transition to pure quark matter. In practice, this leads to a reduced stellar radius at fixed mass $M$, which increases the mean-density scaling $M/R^{3}$ and consequently results in a higher $f$-mode frequency. 

Next, we consider the dashed green and orange lines. These scenarios involve fermionic DM (around 1 GeV) produced inside the NS via a neutron decay anomaly in nucleonic matter \cite{Thakur:2025zhi,Shirke:2024ymc}.  These models are qualitatively similar to ours as a DM component is produced within the hadronic phase. However, in our case, the DM consists of a heavier bosonic S particle with a mass of about $2$~GeV, rather than a fermionic species with a mass around $1$~GeV.
Notably, prior to the transition to quark matter, our model exhibits higher frequencies than the green and orange curves. This is due to the simultaneous presence of hyperons and heavy bosonic S particles, which increases the stellar compactness and consequently raises the frequency. After the onset of pure quark matter, however, the behavior changes, as DM and hyperons are no longer present in the core.
 Finally, the dashed red curve shows the model with the massive WIMP-like fermionic DM, which is accreted rather than produced \cite{Flores:2024hts}. This leads to the highest frequencies since the DM particle is very massive in the order of hundreds GeV and softens the EOS significantly. Overall, our qualitative comparison illustrated in Fig. \ref{fig:f-m compare}, shows how different particles, whether hyperons, various DM candidates, or quark matter, affect the $f$-mode frequencies.}

\begin{figure}[htbp]
  \centering
  \begin{minipage}[b]{0.6\linewidth}
    \centering
    \includegraphics[width=\linewidth]{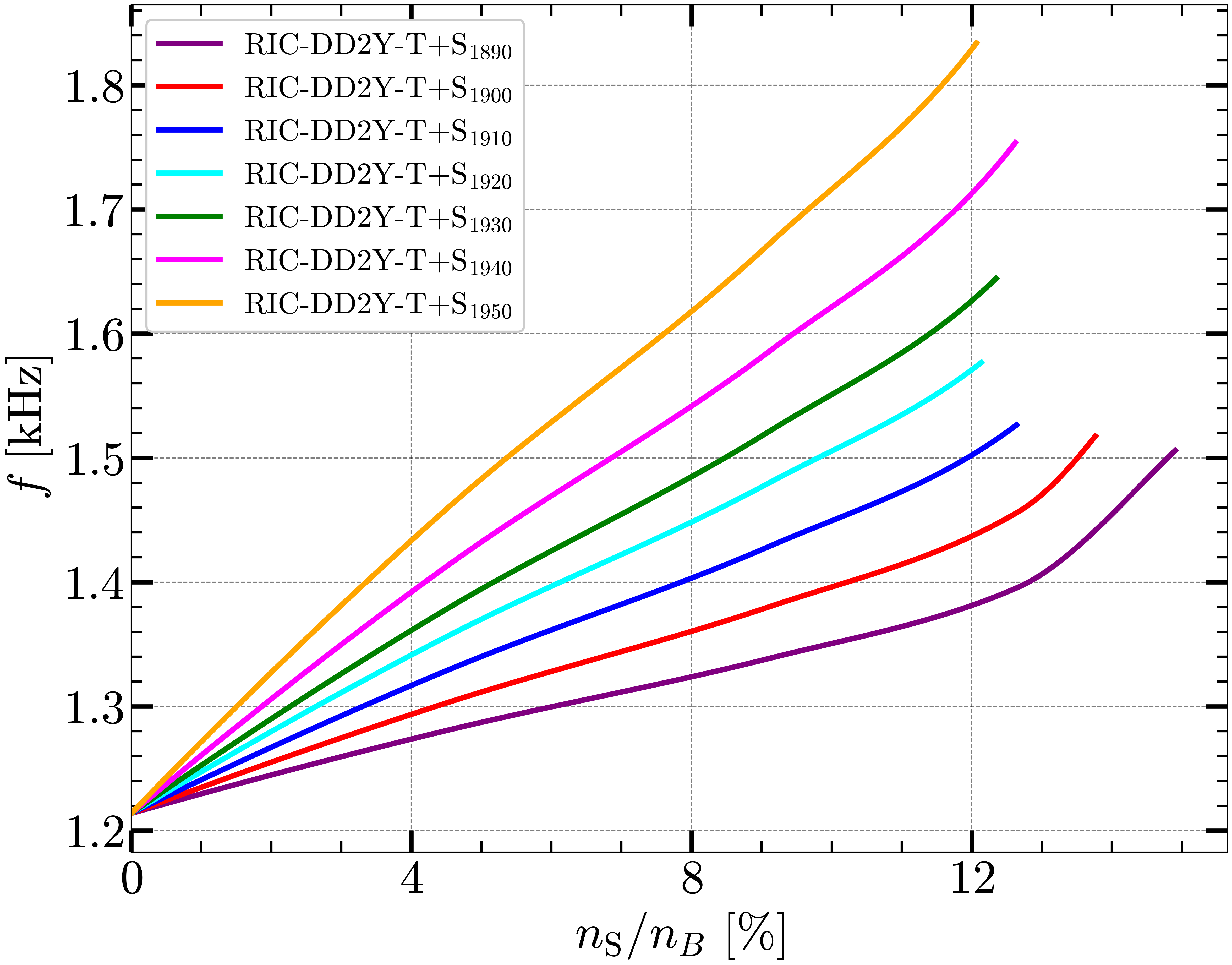}
    \par\medskip
  \end{minipage}
  \caption{{$f$-mode frequency as a function of the S fraction $n_S/n_B$ (in percent), where $n_S$ is the number
density of $S$ particles and $n_B$ is the total baryon density, for stellar configurations before the onset of
phase transition to quark matter. Curves correspond to different S masses $m_S=1890$--$1950$ MeV (see legend).}}
  \label{fig:DM_frac}
\end{figure}

{To quantify how the $f$-mode frequency depends on the DM fraction, we present the $f$-mode frequency as a function of the S fraction $n_S/n_B$ in percentage, where $n_S$ is the number density of S particles and $n_B$ is the total baryon density. Here, $n_S/n_B$ denotes the central fraction of each equilibrium configuration. Each curve is shown up to the onset of the phase transition to quark matter, as beyond this point, the composition changes and the fraction of S is no longer well-defined. It is worth noting that, for the range of S-particle masses considered in this work, the onset of quark matter occurs at S fractions of $12\%$-$15\%$, which represent the maximum DM content attained in the NS core within our model. 

For each $m_S$ in the range $1890$-$1950$ MeV, the $f$-mode frequency increases monotonically with $n_S/n_B$.
 This trend is due to the fact that a larger S fraction softens the hadronic EOS and yields
more compact stars. Moreover, for a fixed DM fraction (e.g., at $12\%$), heavier S masses lead to higher $f$-mode frequencies. This occurs because more massive DM particles make the star more compact by softening the EOS, thus raising the oscillation frequency. Conversely, for a fixed frequency (e.g., around 1.5 kHz), a higher S mass achieves that frequency at a relatively lower DM fraction. This point shows that more massive S-particles influence the star's structure more significantly at lower fractions.}

\begin{figure*}[htbp]
  \centering
  \begin{minipage}[b]{0.32\linewidth}
    \centering
    \includegraphics[width=\linewidth]{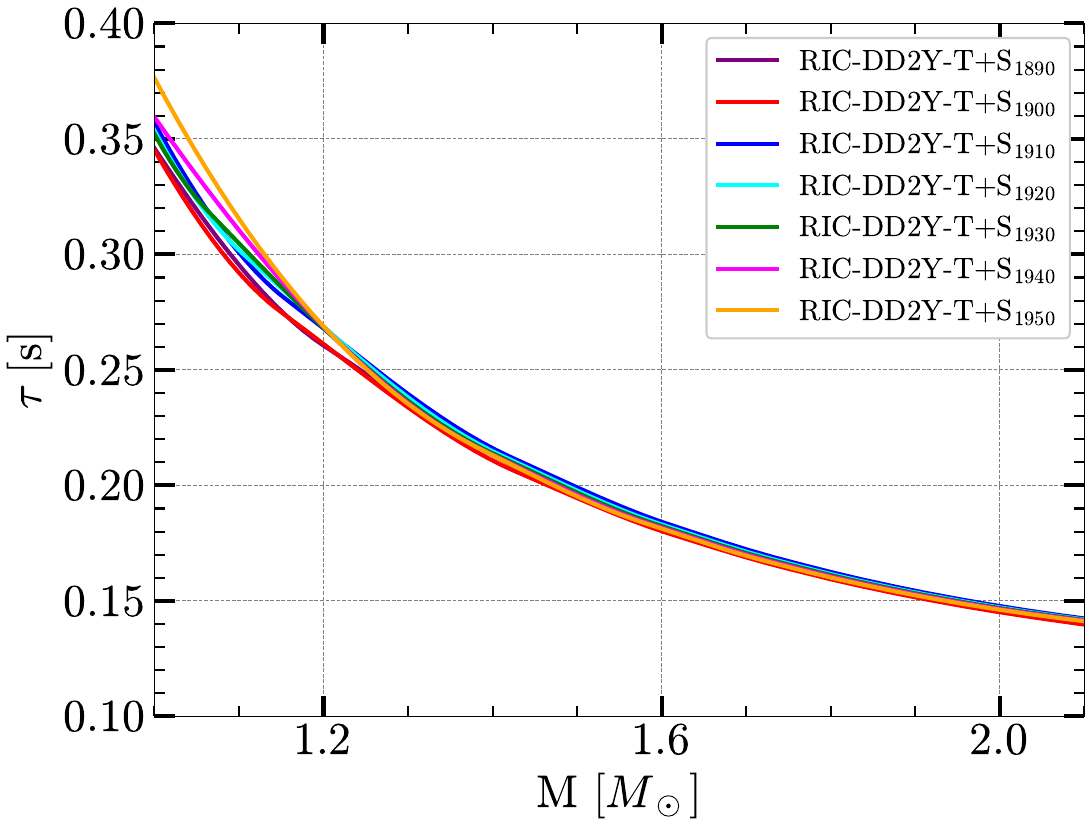}
    \par\medskip
    \centering\textbf{(a)}
  \end{minipage}\hfill
  \begin{minipage}[b]{0.32\linewidth}
    \centering
    \includegraphics[width=\linewidth]{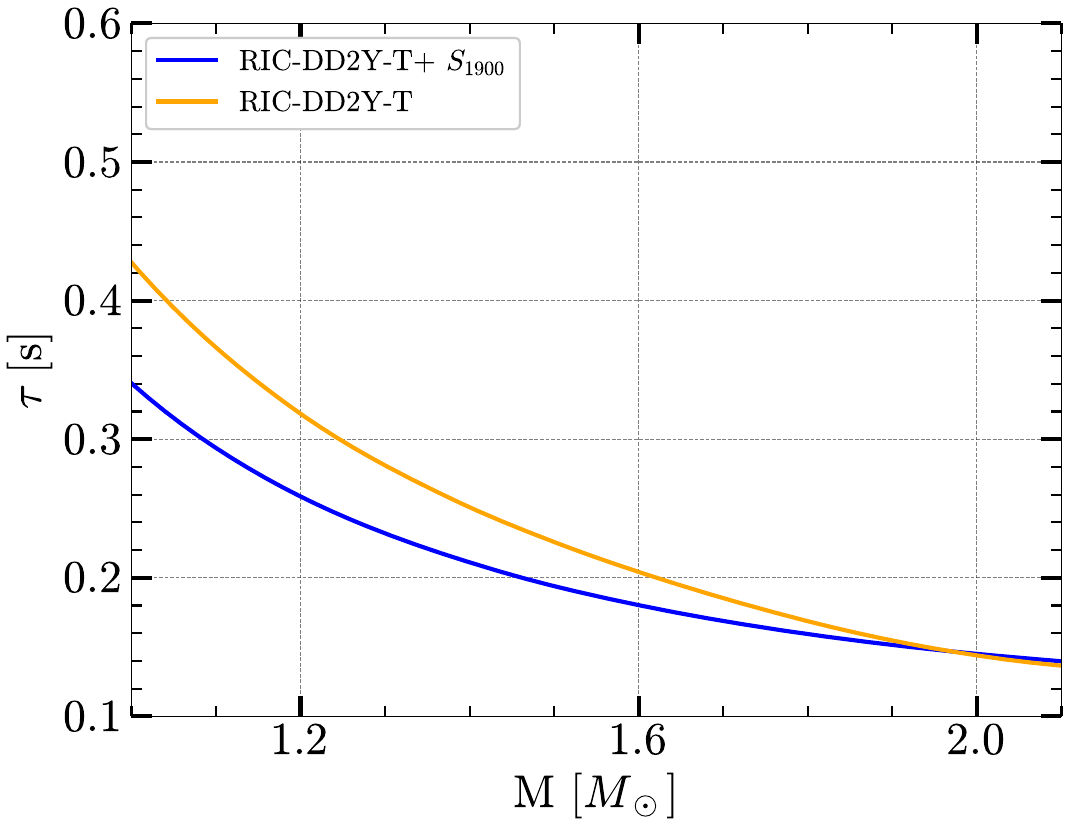}
    \par\medskip
    \centering\textbf{(b)}
  \end{minipage}\hfill
  \begin{minipage}[b]{0.32\linewidth}
    \centering
    \includegraphics[width=\linewidth]{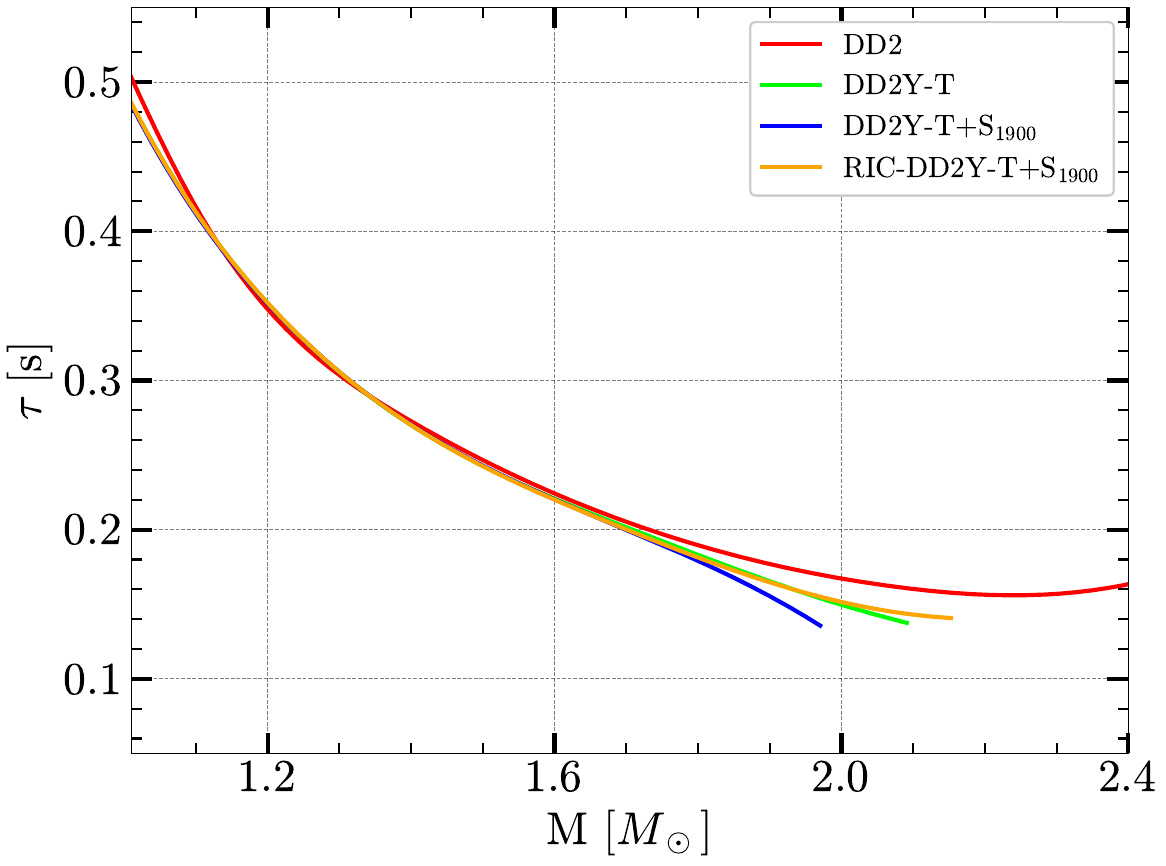}
    \par\medskip
    \centering\textbf{(c)}
  \end{minipage}
  \caption{The $f$-mode frequency damping time ($\tau$), in seconds, plotted as a function of stellar mass. {Panel (a) presents the results for different S masses. Panel (b) compares hybrid stars without DM (orange line) to hybrid stars with DM for $m_s = 1900~\mathrm{MeV}$ and $x = 0.03$ (blue line). Panel (c) shows the damping time for a DM-admixed hybrid star (orange line), alongside the corresponding nucleonic star (red line), hyperonic star (green line), and DM-admixed hyperonic star (blue line) for $m_s = 1900~\mathrm{MeV}$ and $x = 0.08$.}
}
  \label{fig:damping_time_all}
\end{figure*}

Figure~\ref{fig:damping_time_all} displays the damping time $\tau$ (in seconds) of $f$-mode oscillations as a function of gravitational mass $M$ (in $M_\odot$). The damping time $\tau$ characterizes the timescale over which the amplitude of the $f$-mode oscillation decays due to the emission of GWs. For a typical NS, the $f$-mode frequency lies within the range of 1-3~kHz, and the corresponding damping time $\tau_f$ is typically a few tenths of a second~\cite{Kunjipurayil:2022zah, Thakur:2025zhi}. As the NS mass increases, the damping time decreases steadily. This behavior is consistently observed across all three panels of figure~\ref{fig:damping_time_all}. In the left panel, it is evident that the damping time remains nearly identical for all hybrid configurations involving S particles. This suggests that, given an early phase transition and nearly identical stiffness of the QM EOSs, the small variations in the hadronic EOS induced by changing the DM mass in steps of $10$~MeV have little effect on the damping time, which therefore appears largely insensitive to these differences.

The middle panel presents a comparison between hybrid EOSs without DM (orange) and DM-admixed configurations (blue), corresponding to the case shown in panel (b) of figure~\ref{fig:frequency_mass_combined}. For NSs with a stiff EOS (RIC-DD2Y-T), the larger radius results in a lower mean density, which weakens the restoring force behind fluid oscillations, leading to lower $f$-mode frequencies. Additionally, the lower compactness reduces metric perturbations and GW damping, resulting in longer damping times. In contrast, NSs with a soft EOS that includes DM (RIC-DD2Y-T+S$_{1900}$) have smaller radii and higher mean densities. Here, the presence of DM directly contributes to this softening and therefore stronger restoring force and higher $f$-mode frequencies. The increased compactness caused by DM also enhances metric perturbations and shortens the damping time, which can be seen in Table~\ref{tab:hybrid_frequency_damping_comparison} and enables more efficient emission of GWs~\cite{Lindblom:1983ps,PhysRevD.84.044017}.

Finally, we examine the damping time as a function of NS mass ($M_\odot$) for the third scenario, in which we investigate the effect of sequentially adding additional degrees of freedom.
 As expected, the damping time ($\tau$) gradually decreases with increasing mass. From the right panel of figure~\ref{fig:damping_time_all} and Table~\ref{fig:damping_time_all}, it is evident that $\tau$ remains nearly constant up to around 1.8\,$M_\odot$. Near the maximum-mass configuration, the DD2Y-T+S$_{1900}$ EOS, the softest model, containing large fractions of both hyperons and S DM, yields the shortest damping time of 0.135\,s, whereas the stiffest EOS, DD2, shows the longest damping time of 0.161\,s.  

In between, the RIC-DD2Y-T+S$_{1900}$ configuration, which is the only hybrid model in this scenario, competes in damping time with DD2Y-T, the hadronic case containing only nucleons and hyperons. Initially, the phase transition to QM stiffens the hybrid EOS relative to DD2Y-T+S$_{1900}$, though it remains softer than DD2Y-T. Consequently, the damping time for RIC-DD2Y-T+S$_{1900}$ is longer than that of DD2Y-T+S$_{1900}$ but shorter than DD2Y-T. At higher masses, however, the increasing dominance of hyperons softens DD2Y-T relative to the hybrid EOS, causing the damping time of the hybrid configuration to exceed that of DD2Y-T for the same stellar mass.

In addition to global observables such as mass, radius, and compactness, the tidal deformability is a crucial parameter for constraining the NS EOS. During the inspiral phase of a binary neutron star (BNS) system, the strong gravitational interaction induces measurable deformations that are highly sensitive to the internal composition of the stars~\cite{Hinderer:2009ca}. Independent and simultaneous measurements of the $f$-mode frequency and tidal deformability can therefore provide valuable constraints on the EOS and the underlying composition of NSs. We present the variation of $f$-mode frequencies as a function of the dimensionless tidal deformability ($\Lambda$) in figure~\ref{fig:tidal}. Our results for the $f$-mode frequencies fall within the bounds inferred from GW170817, which estimates values between 1.43-2.90~kHz for the more massive NS and between 1.48-3.18~kHz for the less massive one~\cite{Pradhan:2022vdf, Pradhan:2020amo}. Moreover, the analysis of $f$-mode frequencies with respect to tidal deformability indicates that, in panel (a), the frequencies are nearly indistinguishable across all studied EOSs, whereas in panel (c), beyond a tidal deformability parameter of $\Lambda \approx 300$, the $f$-mode frequencies also become effectively indistinguishable. This is consistent with the differences between the EOSs discussed for the first and third scenarios in figures~\ref{fig:frequency_mass_combined} and \ref{fig:damping_time_all}. In the first scenario (panel (a)), varying the mass of S DM in the range $m_S = 1890$-$1950$ MeV does not significantly affect the $f$-mode-tidal deformability relation. In the third scenario (panel (c)), this behavior reflects that, in the low mass and thus, the high-deformability regime, the four different EOSs all originate from the same baseline EOS (DD2) and then progressively diverge as additional degrees of freedom emerge, moving toward higher masses and correspondingly lower tidal deformabilities. Again, the ordering of EOS softening and stiffening in this region determines the behavior of the $f$-mode frequencies as a function of tidal deformability.


In contrast, the frequencies in panel~(b) are clearly distinguishable, highlighting the different behavior of the two extreme EOSs: RIC-DD2Y-T without DM (orange curve) and the DM-admixed configuration (blue curve). This emphasizes how bosonic DM, such as S particles with quark content $uuddss$ and a mass of approximately $1900$ MeV, plays a crucial role in differentiating the $f$-mode frequency behavior and in the analysis of GW signals.
 Our results of Figure~\ref{fig:tidal} are in agreement with previous studies~\cite{Pradhan:2022vdf,Rather:2024mtd,Kunjipurayil:2022zah}.

\begin{figure*}[htbp]
  \centering
  \begin{minipage}[b]{0.32\linewidth}
    \centering
    \includegraphics[width=\linewidth]{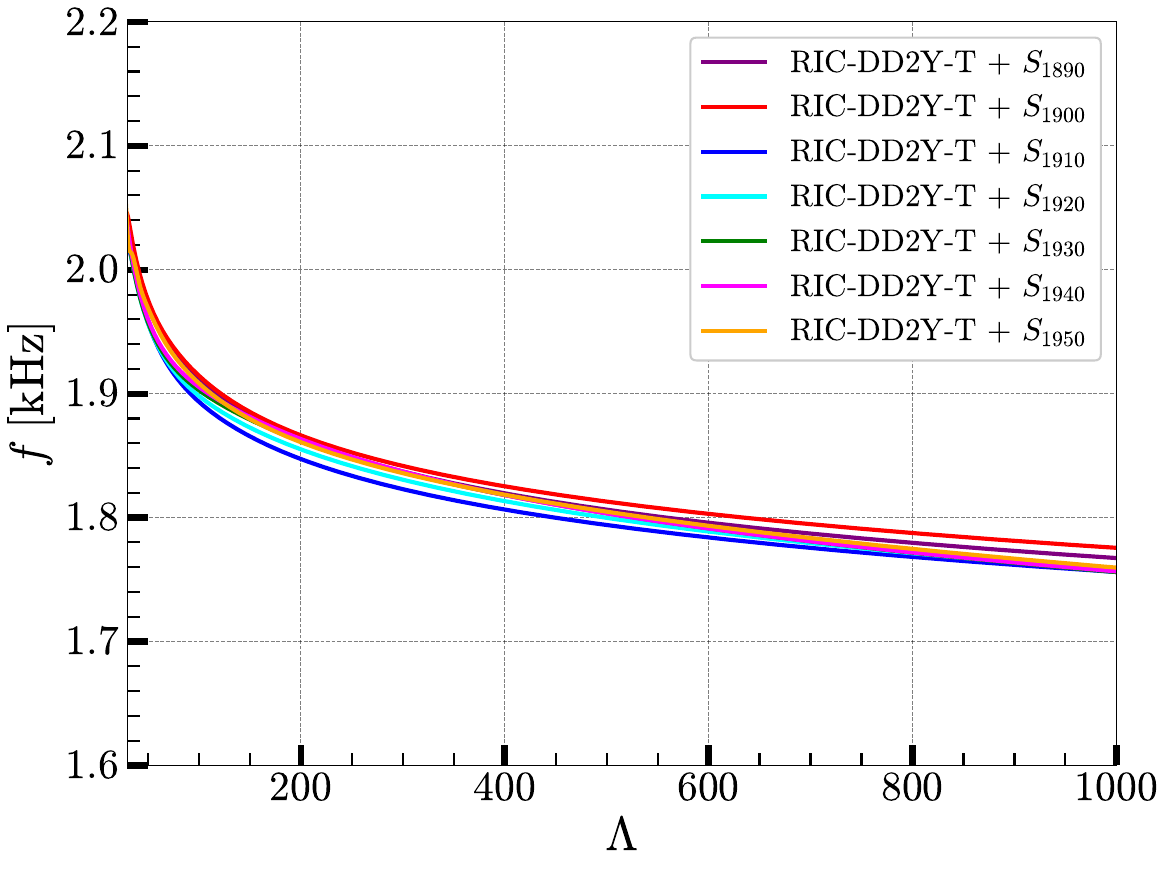}
    \par\medskip
    \centering\textbf{(a)}
  \end{minipage}\hfill
  \begin{minipage}[b]{0.32\linewidth}
    \centering
    \includegraphics[width=\linewidth]{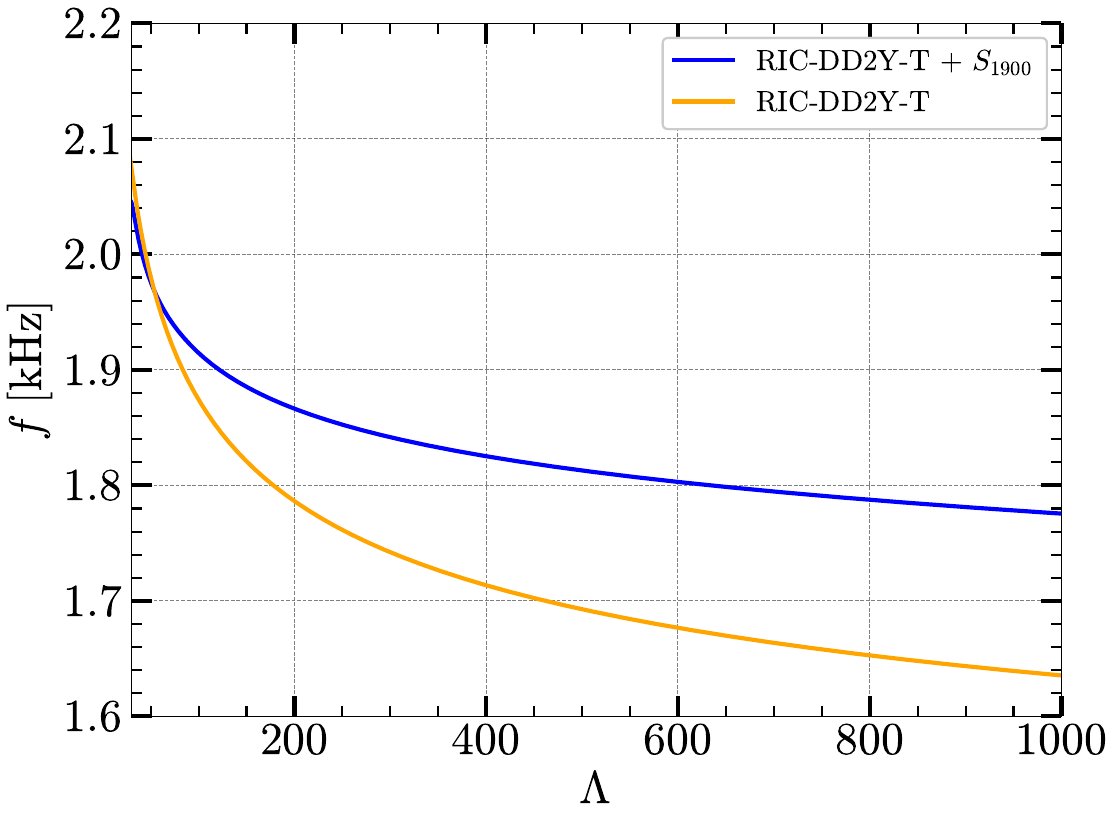}
    \par\medskip
    \centering\textbf{(b)}
  \end{minipage}\hfill
  \begin{minipage}[b]{0.32\linewidth}
    \centering
    \includegraphics[width=\linewidth]{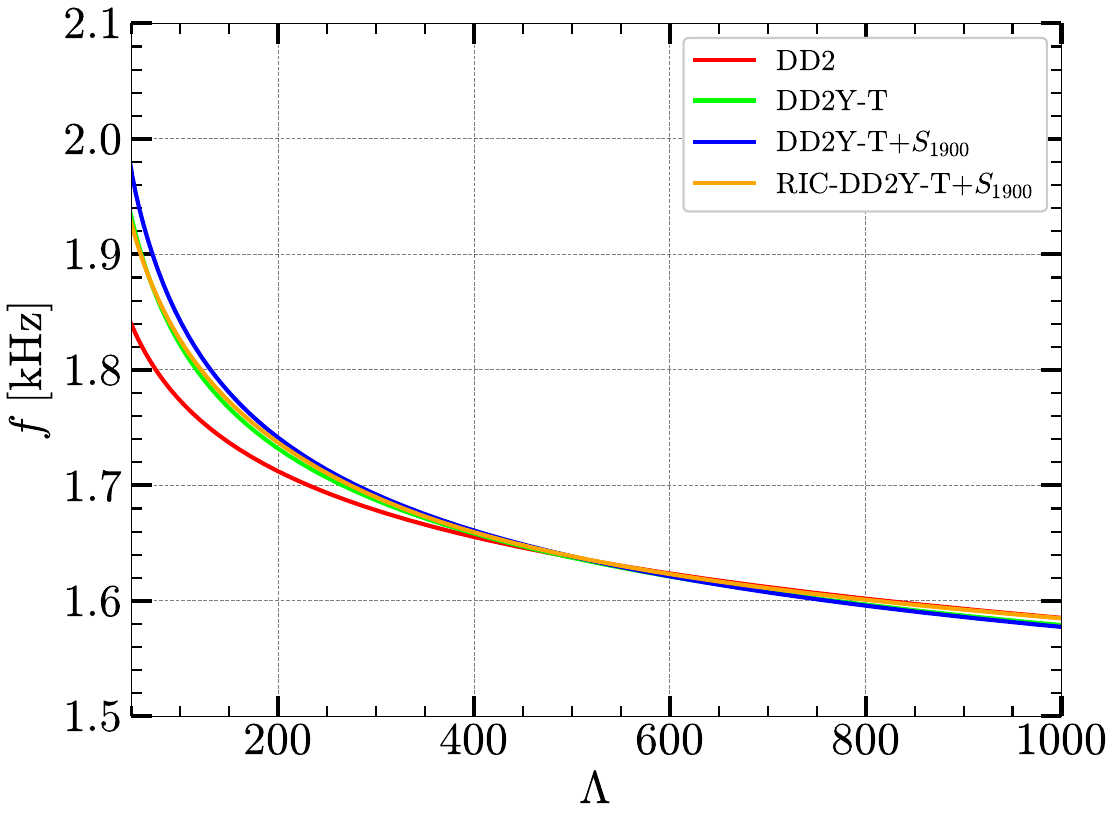}
    \par\medskip
    \centering\textbf{(c)}
  \end{minipage}
   \caption{The $f$-mode frequency (in kHz) as a function of the dimensionless tidal deformability $\Lambda$ for all three scenarios. {Panel (a) shows the results for the different S-mass configurations. Panel (b) contrasts hybrid stars without DM (orange line) with DM-admixed hybrid stars for $m_s = 1900~\mathrm{MeV}$ and $x = 0.03$ (blue line). Panel (c) presents the corresponding results for a DM-admixed hybrid star (orange line), along with the nucleonic star (red line), hyperonic star (green line), and DM-admixed hyperonic star (blue line) for $m_s = 1900~\mathrm{MeV}$ and $x = 0.08$.}}
  \label{fig:tidal}
\end{figure*}

\section{Quasi universal relations}
\label{sec:universal}

Quasi-universal relations, which exhibit minimal dependence on the underlying EOS, play an essential role in NS studies. Directly determining NS properties from GW frequency measurements is difficult because of uncertainties in the dense matter EOS. Over the years, several such relations have been identified, for example, the fundamental $f$-mode frequency expressed as a function of the star’s mean density, its compactness, or the compactness dependence of the spacetime $w$-mode frequency and damping time~\cite{Benhar:1998au,Lau:2009bu,Chan:2014kua,Sotani:2021kiw,Zhao:2022tcw}. Once the GW signal characteristics are measured, these relations provide a pathway to infer key macroscopic parameters, such as mass, radius, and compactness. In turn, combining these macroscopic estimates with additional observations offers an indirect yet powerful means to constrain the EOS of ultra-dense matter.

A widely recognized relation, proposed by \cite{Andersson:1997rn}, describes the connection between the $f$-mode frequency and the star's mean density, expressed as:
\begin{equation}
\frac{f}{\text{kHz}}= a + b \sqrt{\frac{M}{R^3}}.
\label{eq:f_mode_relation}
\end{equation}
where a and b are 0.22 and 32.16 respectively.

\begin{table}[ht]
\centering
\caption{Other fits used in this work for $f$ [kHz] as a function of mean density $\sqrt{M/R^3}$.}
\begin{tabular}{lcc}
\hline
Fit & $a$ & $b$ \\
\hline
SB~\cite{Shirke:2024ymc}   & 0.630 & 33.540 \\
Thakur et al.~\cite{Thakur:2025zhi}   & 0.473 & 36.706 \\
IR1~\cite{Rather:2024mtd}  & 0.440 & 37.900 \\
IR2~\cite{Rather:2024mtd}  & 0.390 & 39.440 \\
\hline
\end{tabular}
\label{tabbUR1}
\end{table}

The upper panel of Figure~\ref{fig:UR1} presents the quasi-universal relation between the $f$-mode frequency and $\sqrt{M/R^3}$ for the hybrid EOSs RIC-DD2Y-T+S, constructed with a crossover phase transition from a hadronic phase containing nucleons, hyperons, and a bosonic DM to a deconfined QM phase; colored curves correspond to different S-particle masses at fixed coupling $x = 0.03$.
The ``Our Fit'' curve (black dotted line) represents a quadratic relation, which is represented by Eq.~\ref{UR1},

\begin{equation}\label{UR1}
\frac{f}{\text{kHz}} = a + b \left( \sqrt{\dfrac{M}{R^3}} \right) + c \left( \sqrt{\dfrac{M}{R^3}} \right)^2 
\end{equation}
where $a = 0.840846$, $b = 18.569058$ and $c = 229.339724$

Above relation is in contrast to the linear fits from other models overplotted for comparison: SB~\cite{Shirke:2024ymc} (based on the neutron-decay DM model), Thakur et al.~\cite{Thakur:2025zhi} (nonlinear RMF, $\sigma$-cut modified RMF, and its neutron decay DM extension), IR1~\cite{Rather:2024mtd} (including hyperons and $\Delta$ baryons without phase transitions), and IR2~\cite{Rather:2024mtd} (which adds a hadron–quark phase transition to the IR1 setup). The fitting coefficients for each model are listed in Table~\ref{tabbUR1}. The choice of a quadratic form in our case is not arbitrary but physically motivated by the intricate microphysics of the underlying hybrid EOS. Specifically, the presence of a smooth crossover to deconfined quark matter, along with the inclusion of a dark sector, introduces significant nonlinearities in the $f$-mode frequency behavior. Our quadratic fit successfully tracks the full sequence across the entire range of mean stellar densities, mitigating the systematic deviations observed in the linear fits, especially at the low and high ends of the density spectrum. This higher-order correction captures essential physical effects such as the gradual formation of a quark core, the emergence of hyperons and the influence of DM mass, all of which modulate the stiffness and sound-speed profile of the star in a nontrivial way—necessitating a departure from strictly linear scaling. Interestingly, the linear fit labeled ``SB'', which incorporates the neutron decay dark matter extension, provides an accurate description of the mid-range mean density regime. The lower panel validates this choice by showing the relative percentage error, $|\Delta f / f|\%$, between our numerical data and the quadratic best-fit line. The error remains remarkably small, peaking at just over 3\% and staying well below 1\% for higher compactness values, which confirms that the quadratic formula provides an excellent and physically motivated description of the $f$-mode behavior for these specific hybrid star models.

\begin{figure}[htbp]
  \centering
  \begin{minipage}[b]{0.6\linewidth}
    \centering
    \includegraphics[width=\linewidth]{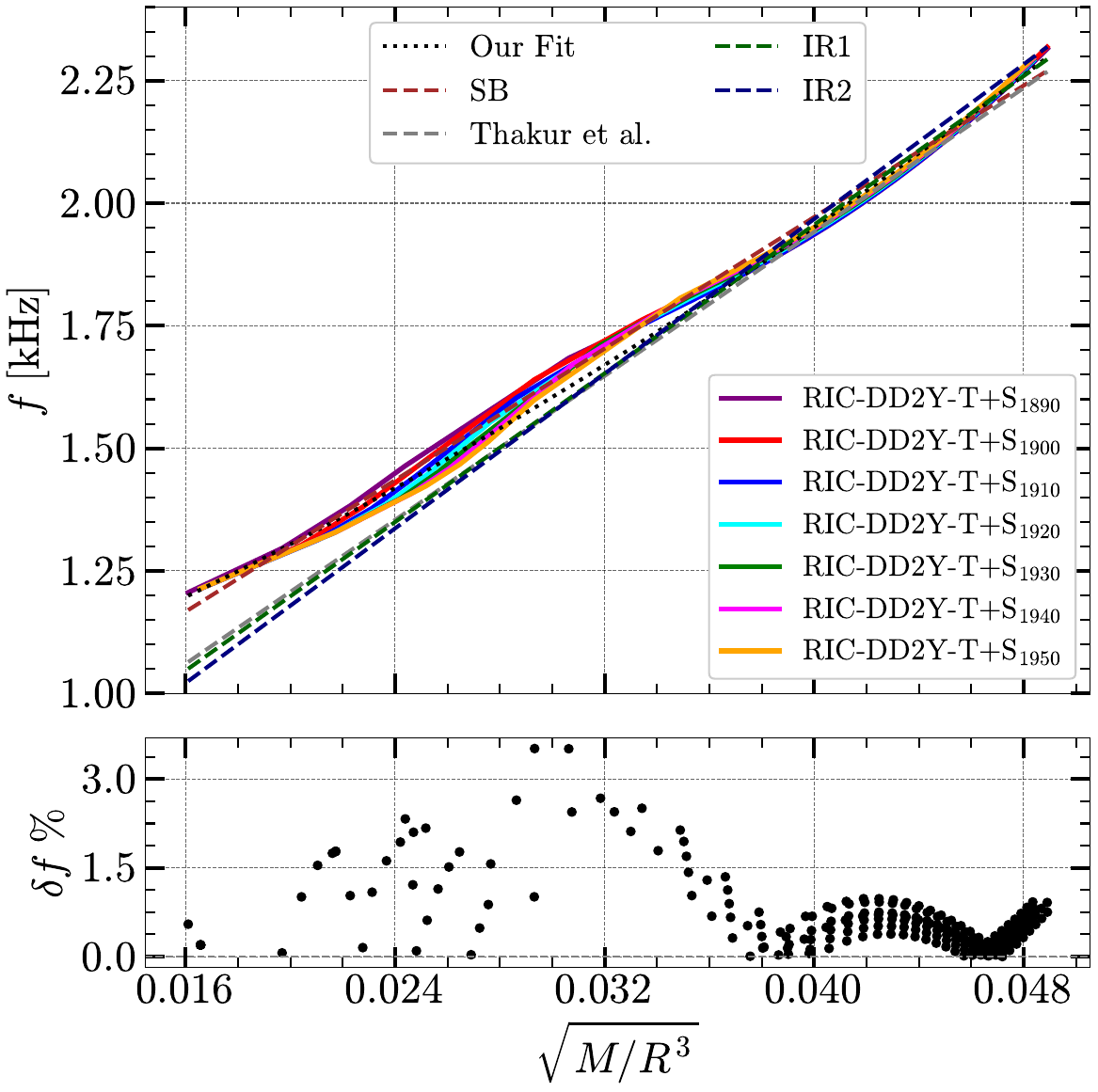}
    \par\medskip
  \end{minipage}
  \caption{The $f$-mode frequency as a function of the mean stellar density $\sqrt{M/R^{3}}$ for the first scenario. {Several empirical fits from previous studies are shown as dashed lines (SB (\cite{Shirke:2024ymc}, Thakur et al. \cite{Thakur:2025zhi}, and IR1, IR2 \cite{Rather:2024mtd}), while the dotted black line represents our best-fitting relation.} The lower panel displays the relative deviation (\%) between the numerical data and our fitted curve.}
  \label{fig:UR1}
\end{figure}

\begin{figure}[htbp]
  \centering
  \begin{minipage}[b]{0.6\linewidth}
    \centering
    \includegraphics[width=\linewidth]{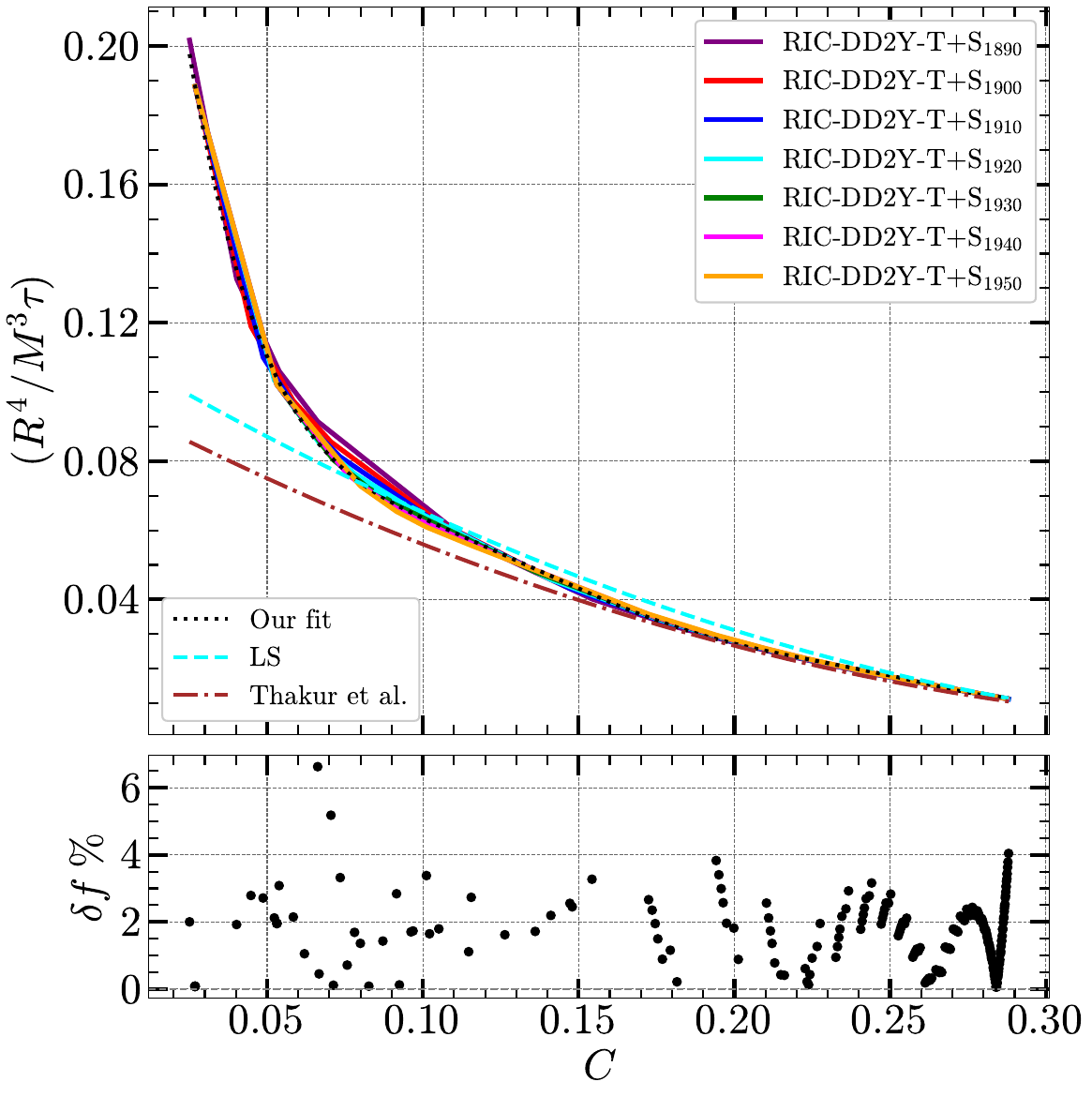}
    \par\medskip
  \end{minipage}
  \caption{The normalized damping time $R^{4}/(M^{3}\tau)$ of the $f$-mode as a function of the stellar compactness $C = M/R$ for the first scenario. {The cyan and brown curves correspond to the fits of Ref.~\cite{Lioutas:2017xtn} and Ref.~\cite{Thakur:2025zhi}, respectively, while the black dotted line denotes our best-fitting relation.} The lower panel shows the percentage relative error between the numerical data and our fitted curve.}
  \label{fig:UR2}
\end{figure}

Andersson and Kokkotas~\cite{Andersson:1997rn} first introduced a quasi-universal relation for the $f$-mode scaled damping time, $\tau$, which was subsequently refined by Lioutas and Stergioulas (LS)~\cite{Lioutas:2017xtn} through the inclusion of higher-order terms as shown in table ~\ref{tabbUR2}. Building upon these studies, Figure~\ref{fig:UR2} shows the quasi-universal relation between the scaled damping-time variable $\left( R^4 / M^3 \tau \right)$ and the compactness $C = M / R$ for the same hybrid EOSs RIC-DD2Y-T+S. All sequences follow a common, smooth monotonic decrease with compactness $C$: the slope is steep at low compactness and progressively flattens as $C$ increases.  We compare our best fit (black dotted), which adopts a higher-order polynomial in $C$ represented by below equation Eq.~\eqref{UR2} with other fit relations, notably the LS relation (cyan)~\cite{Lioutas:2017xtn}, which is based primarily on microscopic nuclear models, and the relation proposed by Thakur et al.\ (brown)~\cite{Thakur:2025zhi}, which incorporates nonlinear RMF models, the $\sigma$-cut modified RMF, and its dark matter-extended neutron decay version.

\begin{equation}\label{UR2}
\frac{R^4}{M^3 \tau} = a + b\,C + c\,C^2 + d\,C^3 + e\,C^4 + f\,C^5 + g\,C^6
\end{equation}
where, $a = 0.4095642$, $b = -12.26236$, $c = 186.0756$, $d = -1505.978$, $e = 6603.648$, $f = -14837.04$ and $g = 13389.98$

\begin{table}[ht]

\centering
\caption{The other fitting relations used in this study represent $R^4/(M^3\tau)$ as a function of the compactness $M/R$.}
\begin{tabular}{lccc}
\hline
Fit & $a$ & $b$ & $c$ \\
\hline
LS~\cite{Lioutas:2017xtn} & 0.112 & -0.53  & 0.628 \\
Thakur et al.~\cite{Thakur:2025zhi} & 0.097 & -0.469 & 0.586 \\
\hline
\end{tabular}
\label{tabbUR2}
\end{table}

While the quadratic forms reproduce the mid-range behavior, they systematically overestimate $\left( R^4 / M^3 \tau \right)$ at low $C$ and underestimate it at intermediate–high $C$, failing to capture the pronounced curvature and change of slope associated with composition-driven microphysics and with strong-gravity effects as $C$ increases. The added terms in our polynomial provide the minimal smooth flexibility needed to track these density-dependent departures from simple scaling across the full compactness range, yielding visibly improved agreement from the least to the most compact models. The lower panel quantifies the success of this complex fit, showing that the relative percentage error $|\delta f |$ remains mostly below 4\%, validating that the sixth-order polynomial is a robust representation of this nuanced physical relationship, which simpler models would fail to describe accurately.

\begin{figure}[htbp]
  \centering
  \begin{minipage}[b]{0.6\linewidth}
    \centering
    \includegraphics[width=\linewidth]{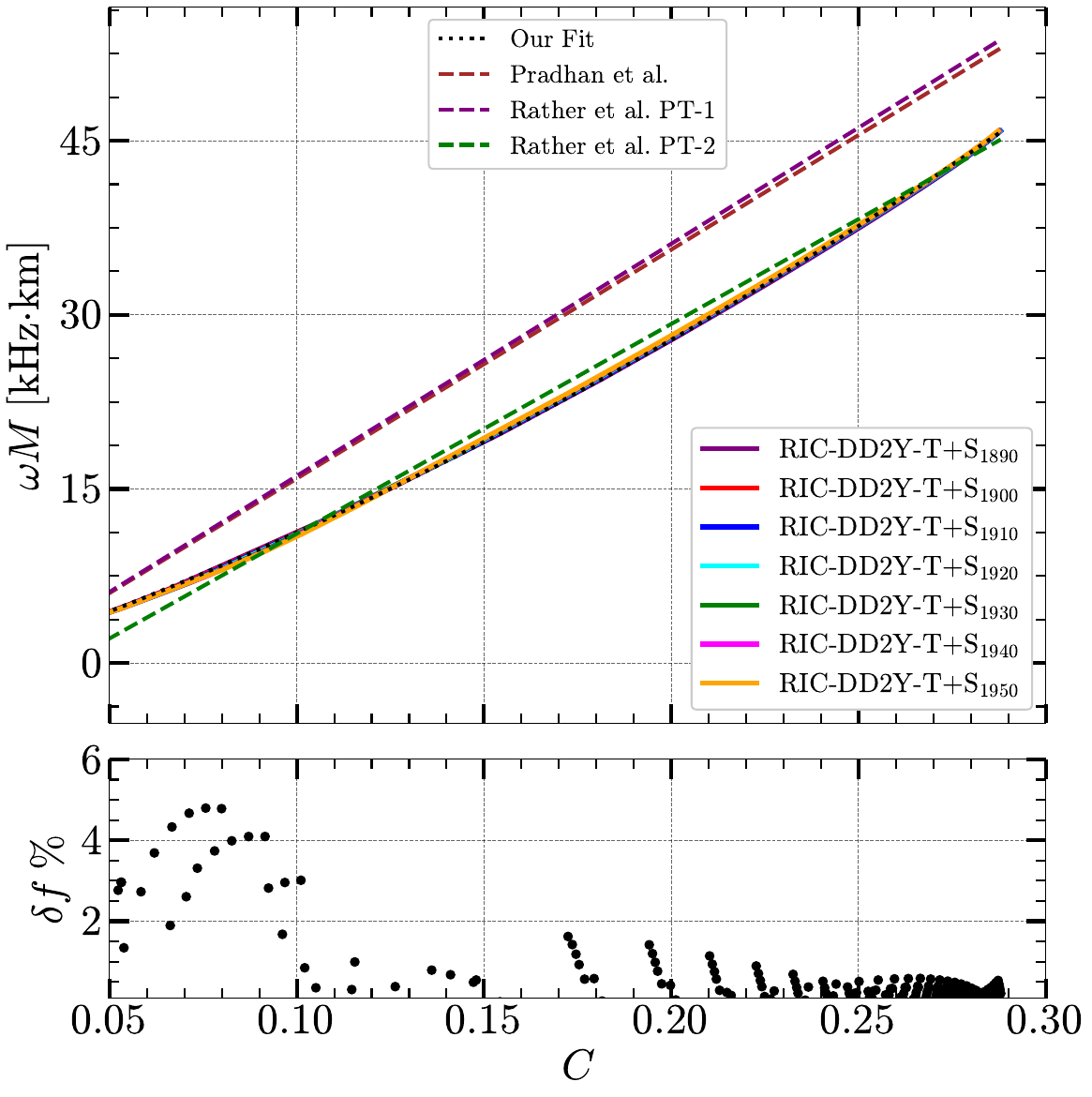}
    \par\medskip
  \end{minipage}
  \caption{The angular frequency $\omega = 2\pi f$, scaled by the stellar mass ($\omega M$), as a function of compactness for the first scenario. {The red dashed line corresponds to the fit by Pradhan et al.~\cite{Pradhan:2020amo}, while the purple and green dashed curves show the PT-1 and PT-2 parameterizations of Rather et al.~\cite{Rather:2024mtd}. The black dotted line represents our best-fitting relation.} The lower panel displays the percentage relative error between the numerical data and our fitted curve.}
  \label{fig:UR3}
\end{figure}

Figure~\ref{fig:UR3} presents another important universal relation, illustrating the dimensionless angular frequency of f mode, \( \omega M \) (with \( \omega = 2\pi f \)), as a function of the stellar compactness \( C = M/R \) for the hybrid EOSs RIC-DD2Y-T+S, whose properties have been discussed earlier. The sequences exhibit a smooth, monotonic increase with \( C \), accompanied by a mild positive curvature at higher compactness values. Our best-fit (black dotted), based on the quadratic form given in Eq.~\eqref{UR3}, accurately reproduces the model trends across the entire compactness range. 

\begin{equation}\label{UR3}
\omega M = a\,C^{2} + b\,C + c
\end{equation}
where a = 199.734977, b = 106.082697, c = - 1.365759
and have units of kHz$\cdot$km

For comparison, we also show linear fits from previous studies, with their coefficients summarized in Table~\ref{tabbUR3}. The red dashed curve corresponds to the fit by Pradhan et al.~\cite{Pradhan:2020amo}, which includes nucleons and hyperons and is based on the Cowling approximation for the $f$-mode calculation. The purple and green dashed curves represent the parameterizations Rather et al. PT-1~\cite{Rather:2024mtd} and Rather et al. PT-2~\cite{Rather:2024mtd}, respectively, both models include hyperons and $\Delta$ baryons with a phase transition. Notably, the fits by Pradhan et al. and Rather et al. PT-1 show significant deviation from our result. In contrast, Rather et al. PT-2 lies closer to our fit and captures some regions well, though it still exhibits notable discrepancies, particularly in the low and intermediate compactness range. The inclusion of a quadratic term in our fit is physically motivated; it accounts for composition-dependent variations in the effective stiffness of the star, which naturally introduce curvature in the \( \omega M \)--\( C \) relation, deviating from a purely linear behavior.

\begin{table}[ht]

\centering
\caption{Other fits from references for $\omega M$ [kHz$\cdot$km] as a function of compactness $C$.}
\begin{tabular}{lcc}
\hline
Fit & $a$ & $b$ \\
\hline
Pradhan et al.~\cite{Pradhan:2020amo}     & 197.30 & $-3.84$ \\
Rather et al. PT-1~\cite{Rather:2024mtd} & 200.00 & $-3.88$ \\
Rather et al. PT-2~\cite{Rather:2024mtd} & 180.65 & $-6.92$ \\
\hline
\end{tabular}
\label{tabbUR3}
\end{table}
The lower panel validates this approach by showing the relative error (\( \delta f \%\)) of the quadratic fit. The error is largest (about 4--5\%) at low compactness, precisely where the physical effects of the crossover are most significant and introduce the strongest curvature. However, the error decreases rapidly and becomes exceptionally small (below 2\%) for more compact stars. This confirms that the quadratic model is not only highly accurate but also physically essential to capture the oscillation properties of these complex hybrid stars.

\begin{figure}[htbp]
  \centering
  \begin{minipage}[b]{0.6\linewidth}
    \centering
    \includegraphics[width=\linewidth]{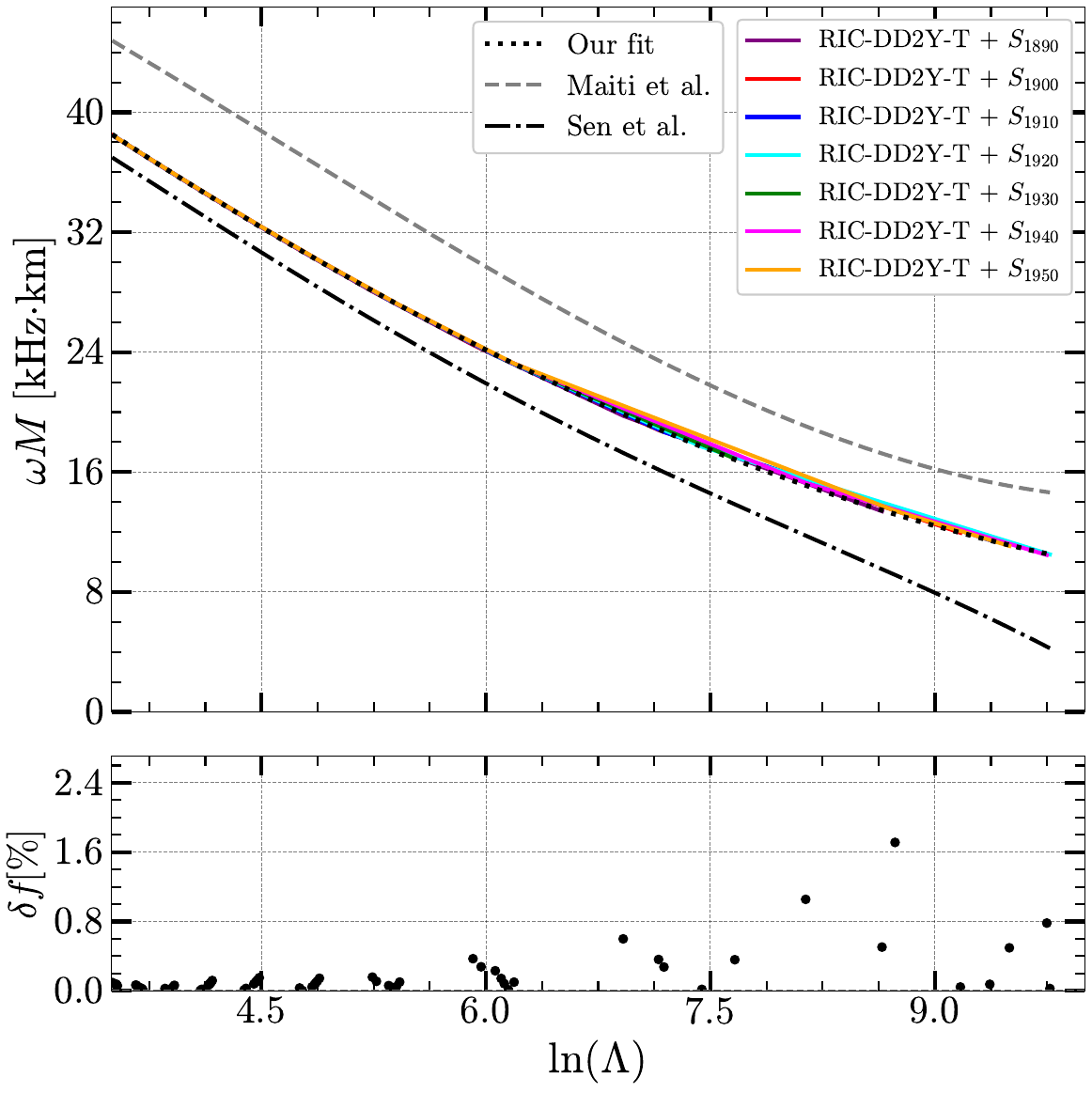}
    \par\medskip
  \end{minipage}
  \caption{The angular frequency ($\omega = 2\pi f$) scaled by the mass ($\omega M$) as a function of the dimensionless tidal deformability ($\Lambda$). 
{The grey dashed line represents the fit proposed by Maiti \textit{et al.}~\cite{Maiti:2025drj}, 
while the black dash-dotted line corresponds to the fit of Sen \textit{et al.}~\cite{Sen:2025ndg}.} 
The lower panel displays the relative error (\%) between the data and our best-fit relation.}
  \label{fig:UR4}
\end{figure}

Figure~\ref{fig:UR4} illustrates the variation of the scaled frequency, $\omega M$ (in kHz$\cdot$km), as a function of the dimensionless tidal deformability parameter $\Lambda$, presented in terms of $\ln\Lambda$. The upper panel compares the present (our) fit corresponding to the first scenario with those proposed by Maiti et al.\ and Sen et al. The lower panel displays the percentage deviation ($\delta f[\%]$). The solid-colored curves in the upper plot represent the different EOS models used, all showing excellent consistency and lying nearly indistinguishable from each other, indicating a robust quasi-universality in the $\omega M$–$\ln\Lambda$ relation. The dotted line denotes our newly proposed fit, given by:

\begin{equation}
\omega M = a + b \, (\ln \Lambda) + c \, (\ln \Lambda)^{2} + d \, (\ln \Lambda)^{3}
\end{equation}

where $a = 67.426668$, $b = -9.797683$, $c = 0.470981$, and $d = -0.006726$, which exhibits excellent agreement with the numerical results across the full range of $\ln\Lambda$. In contrast, the dashed and dash-dotted lines correspond to the analytical relations of Maiti et al.\ \cite{Maiti:2025drj} and Sen et al. \cite{Sen:2025ndg}, respectively. The values of the coefficients are presented in table~\ref{tabbUR4}

\begin{table}[ht]
\centering
\caption{Other fits from references for $\omega M$ [kHz$\cdot$km] as a function of $\ln(\Lambda$).}
\begin{tabular}{lccccc}
\hline
Fit & $a$ & $b$ & $c$ & $d$ & $e$ \\
\hline
Maiti et al.~\cite{Maiti:2025drj} & 60.48 & $-2.31$ & $-0.83$ & $0.06$ & -- \\
Sen et al.~\cite{Sen:2025ndg} & 52.854 & $-0.494$ & $-1.899$ & $0.25$ & $-0.0105$ \\
\hline
\end{tabular}
\label{tabbUR4}
\end{table}

The Maiti et al. systematically overestimates $\omega M$ at entire range of $\ln\Lambda$ values, while the Sen et al. underestimates it, particularly in the high-compactness regime. It is important to emphasize that the model proposed by Maiti et al. incorporates hyperons and H-dibaryons within neutron stars, whereas the model developed by Sen et al. describes dark matter–admixed strange quark stars. The lower panel shows that our fit yields residuals well within $\pm2\%$, confirming its accuracy and improved predictive capability compared to previous fits. Overall, the figure demonstrates that the present relation captures the nonlinear dependence of $\omega M$ on $\ln\Lambda$ more effectively than earlier parametrizations, providing an accurate, description across different EoS models.

\section{Conclusions and Summary} 
\label{sec:summary}

In this work, we have explored the impact of a bosonic DM candidate, S, together with hyperonic degrees of freedom and deconfined QM, on the fundamental $f$-mode oscillations of NSs. By embedding nucleonic, hyperonic, and DM components in a unified hadronic EOS, and modeling the phase transition to QM via a smooth crossover, we constructed hybrid star models that satisfy all current astrophysical constraints on mass, radius, and tidal deformability, including those inferred from GW170817 and the most recent NICER mass-radius measurements for PSR J0740+6620, PSR J0030+0451, PSR J0437-4715, and PSR J0614-3329. Within this framework, we performed fully general relativistic calculations of $f$-mode frequencies and damping times, providing insight into how exotic components may leave their imprint on the GW emission of NSs.

We examined three complementary scenarios. First, varying the S mass at fixed small coupling revealed that lighter DM candidates produce smaller stellar radii and higher $f$-mode frequencies. Second, contrasting hybrid stars with and without DM demonstrated that the presence of bosonic DM modifies stellar compactness, raising oscillation frequencies at low masses where the hadronic EOS is dominated. Third, systematically adding hyperons, DM, and QM to nucleonic matter showed that the interplay among these degrees of freedom reshapes both stellar structure and oscillation spectra. In particular, the competition between hyperons and DM governs the emergence of strangeness: light S DM with weak coupling suppresses hyperons, whereas heavier or more strongly coupled S DM allows their reappearance at higher densities.

Our analysis of damping times further established that DM generally shortens damping times by increasing compactness and enhancing GW emission. Across all scenarios, we found that S DM can substantially affect $f$-mode properties at low and intermediate stellar masses, while at higher densities the dynamics are dominated by hyperons and deconfined QM. These findings highlight the sensitivity of $f$-mode oscillations to the microscopic composition of NS cores and underscore the role of exotic matter in shaping observable GW signals.

We also investigated quasi-universal relations linking $f$-mode characteristics to bulk stellar properties such as mean density, compactness, and tidal deformability. Although simple linear relations break down in the simultaneous presence of hyperons, S DM, and QM, higher-order polynomial fits reproduce the relations with high precision, deviating by only a few percent. While quadratic forms remain adequate for $f$–$\sqrt{M/R^{3}}$ and $\omega M(C)$, the damping-time relation $(R^{4}/M^{3}\tau)(C)$ requires higher-order corrections to capture its curvature properly. In contrast, a cubic fit is sufficient for $f(\Lambda)$. This robustness demonstrates the potential of GW asteroseismology as an EOS-insensitive diagnostic, even for stars containing multiple exotic degrees of freedom.

In summary, our results establish $f$-mode oscillations as a powerful probe of both DM and exotic hyperonic and quark components in NSs. The combined influence of nucleons, hyperons, S DM, and QM leaves distinctive imprints on oscillation frequencies and damping times, which could enable GW observations to distinguish among competing EOS scenarios. Although current detectors lack the sensitivity to resolve $f$-mode signals directly, next-generation observatories such as the Einstein Telescope and Cosmic Explorer are expected to reach the required precision. Their observations would not only provide stringent constraints on the physics of dense matter but also open a unique window onto the fundamental properties of DM.

\section{Acknowledgements}
M. Sh. is supported by NCN under SONATINA 7 grant NO. 2023/48/C/ST2/00297.
D.R.K is supported by SONATINA 7 grant NO. 2023/48/C/ST2/00297. D.R.K is, in part,  supported by Polish NCN Grant No. 2023/51/B/ST9/02798. P.~Thakur is supported by the National Research Foundation of Korea (NRF) grant funded by the Korea government (MSIT) (No.~RS-2024-00457037). This work was supported (in part) by the Yonsei University Research Fund(Yonsei University Frontier Fellowship for Postdoctoral Researchers) of 2025.  M. Sh. and
D. R. K thank the European COST Action CA24101 FuSe for supporting their work.

 \bibliographystyle{JHEP}

\bibliography{references.bib}




\end{document}